\documentclass[prc,showpacs,amsmath,amssymb,superscriptaddress,floatfix,nofootinbib,10pt]{revtex4}

\usepackage{amstext,amsfonts}
\usepackage{mathrsfs}
\usepackage{bbm,bm}
\usepackage{slashed}

\usepackage{graphicx}
\usepackage{caption}
\usepackage{subfig}
\usepackage{booktabs}
\usepackage{array}

\newcommand{\trace}[1]{\ensuremath{\langle #1\rangle}}

\begin{document}
\title{Two-pion exchange contributions to the nucleon-nucleon {interaction} in covariant baryon chiral perturbation theory}
\author{Yang Xiao}
\affiliation{School of Physics, Beihang University, Beijing 100191, China
\&Universite  Paris-Saclay, CNRS/IN2P3, IJCLab, 91405 Orsay, France}

\author{Chun-Xuan Wang}
\affiliation{School of Physics, Beihang University, Beijing 100191, China}

\author{Jun-Xu Lu}
\affiliation{School of Physics, Beihang University, Beijing 100191, China}

\author{Li-Sheng Geng}
\email[E-mail: ]{lisheng.geng@buaa.edu.cn}
\affiliation{School of Physics \& Beijing Advanced Innovation Center for Big Data-based\&
Precision Medicine, Beihang University, Beijing100191, China. \&School of Physics and Microelectronics, Zhengzhou University, Zhengzhou, Henan 450001, China}

\begin{abstract}
Employing the covariant baryon chiral perturbation theory, we calculate the leading and next-to-leading order two-pion exchange (TPE) contributions to $NN$ {interaction} up to order $O(p^3)$. We compare the so-obtained $NN$ phase shifts with $2\leq L\leq 6$ and mixing angles with $2\leq J\leq6$ with those obtained in the nonrelativistic baryon chiral perturbation theory, which allows us to check the relativistic corrections to the {medium}-range part of $NN$ interactions. We show that the contributions of relativistic TPE are more moderate than those of the nonrelativistic TPE. The relativistic corrections play an important role in F-waves especially  the $^3\text{F}_2$ partial wave. Moreover, the relativistic results seem to converge faster than the nonrelativistic results in almost all the partial waves studied in the present work, consistent with the studies performed in the one-baryon sector.
\end{abstract}

\maketitle

\section{Introduction}\label{sec:introduction}
The nucleon-nucleon ($NN$) interaction is one of the most important inputs in nuclear physics and nuclear astrophysics.
Since the seminal work of Yukawa~\cite{ Yukawa:1935xg}, a variety of formulations of $NN$ {interaction} have been proposed and thoroughly studied. Nowadays, a number of  formulations of $NN$ {interaction}, both phenomenological and more microscopic, are already of high precision, in the sense that they can describe $NN$ scattering data with $T_\mathrm{lab}<350$ MeV with a $\chi^2/\mathrm{d.o.f.}\approx 1$. In the phenomenological group, an accurate description of $NN$ scattering data has been achieved  by the Reid93~\cite{Stoks:1994wp}, Argonne $V_{18}$~\cite{Wiringa:1994wb}, and (CD-)Bonn potentials~\cite{Machleidt:2000ge}. Although these phenomenological potentials work very well in describing the $NN$ scattering data, there is no strong connection between these interactions and the underlying theory of the strong interaction, Quantum Chromodynamics (QCD). As for the microscopic ones, chiral effective field theory has achieved astonishing success. In the 1990s, Weinberg proposed that one can construct $NN$ {interaction} using the Heavy Baryon (HB) chiral effective field theory (ChEFT)~\cite{Weinberg:1990rz,Weinberg:1991um}. Following this idea, numerous studies have been performed and the description of $NN$ scattering data has become comparable to the  phenomenological forces since 2003~\cite{Entem:2003ft,Epelbaum:2004fk,Epelbaum:2008ga,Machleidt:2011zz, Epelbaum:2014sza,Entem:2015xwa,Entem:2017gor}. In chiral $NN$  {interaction}, the low energy constants (LECs) responsible for the short-range part of the $NN$
 interaction play an important role for the description of  partial waves with $L\leq 2$ while
 two-pion and one-pion exchanges responsible for the {medium} and long-range parts, respectively, almost saturate  higher partial waves~\cite{Kaiser:1997mw}.  Nonetheless, it was shown in Ref.~\cite{Kaiser:1997mw} that a non-negligible discrepancy between chiral $NN$ phase shifts and the {Nijmegen} partial wave analysis can still be observed in F-waves especially  the $^3\text{F}_2$ and $^3\text{F}_4$ partial waves.

 In a recent work~\cite{Ren:2016jna}, a  study of $NN$  {interaction}  in covariant baryon chiral effective field theory was proposed. Interestingly, it was shown that the leading order (LO) relativistic $NN$ interaction can describe the $NN$ phase shifts as well as the next-to-leading order (NLO) nonrelativistic $NN$ interaction. Of course, there is no mystery here because the covariant formulation can be viewed as a more efficient ordering of chiral expansion series. This can be seen from the fact that at LO the covariant formulation has four LECs~\footnote{In fact, there are five LECs in Ref.~\cite{Ren:2016jna}, but according to the power counting of Ref.~\cite{Xiao:2018jot}, there should be four, both of which provide very similar descriptions of the $J=0$ and 1 partial wave phase shifts~\cite{Wang:2020myr}.} which is in between those of the LO and NLO nonrelativistic ChEFT, two and nine, respectively. It remains to be seen whether relativistic effects or corrections are important in the two-pion exchange contributions. As there are no unknown LECs involved in these contributions, it should be more appropriate to check the relevance of relativistic corrections. More specifically, it would be interesting to  investigate whether the F partial waves can be better described compared to Ref.~\cite{Kaiser:1997mw}. Although in Ref.~\cite{Kaiser:1997mw} a covariant calculation has been performed already, but the potentials are then expanded and only contributions up to the third order are kept. The corrections of higher order were neglected. We prefer not to do the nonrelativistic expansion, so that all relativistic corrections are properly kept to maintain {Lorentz}  invariance.

The main purpose of this work is to study the {medium}-range part of $NN$ {interaction}, or more specifically, the two-pion exchange contributions in covariant baryon chiral effective field theory. In this work, we start from the covariant chiral $\pi N$ Lagrangians up to the second order~\cite{Fettes:2000gb} and construct the relativistic TPE $T$-matrix up to the third order of chiral expansion. All power counting breaking (PCB) terms are removed in the spirit of the extended-on-mass-shell scheme (EOMS)~\cite{Gegelia:1999gf,Fuchs:2003qc}, which has been well established in the one-baryon sector (see Ref.~\cite{Geng:2013xn} for a short review).  Then we compute the $NN$ phase shifts and mixing angles with $L\geq2$ and $J\geq2$.

The paper is organized as follows. In Sec.II, the chiral Lagrangians needed for computing the two-pion exchange contributions are briefly discussed. The TPE $T$-matrix up to the third order is presented in Sec.III. In sec.IV, we compare the so-obtained $NN$ phase shifts with the Nijmegen partial wave analysis and those of Ref.~\cite{Kaiser:1997mw}. A short summary and outlook is given in the last section.

\section{Chiral Lagrangian}\label{sec:lagrangians}

First, we briefly explain the power counting rule in constructing the covariant baryon chiral Lagrangians, for more details, see, e.g., Refs.~\cite{Gegelia:1999gf,Fuchs:2003qc}. The core of an effective field theory lies in the power counting rule, which emphasises the importance of certain Feynman diagrams for a given process. In this work we adopt the naive dimensional analysis, in which amplitudes are expanded in powers of $(p/\Lambda_\chi)$, where $p$ refers to the low energy scale including the three momentum of nucleons and the pion mass, and $\Lambda_\chi$ refers to the chiral symmetry breaking scale. The chiral order $\nu$ of a Feynman diagram (after proper {renormalization}) with $L$ loops is defined as,
\begin{equation}\label{eq:pc}
\nu= 4L - 2N_\pi - N_n +\sum_k k V_k,
\end{equation}
where $N_{\pi, n}$ refers to the number of pion and nucleon propagators, and $V_k$ {denotes} the number of $k$-th order vertices.
It was realized very early that such a definition is not full-filled in the one-baryon sector, because  the large non-zero baryon mass at the chiral limit leads to the so-called power counting breaking problem~\cite{Gasser:1987rb}. Many approaches have been proposed to recover the power counting defined in Eq.~(\ref{eq:pc}), and the most studied ones are the heavy baryon formulation~\cite{Jenkins:1990jv,Bernard:1992qa}, the infrared approach~\cite{Becher:1999he}, and the EOMS approach~\cite{Gegelia:1999gf,Fuchs:2003qc}. In the present work, we adopt the EOMS approach. The exact procedure of removing the PCB terms in the EOMS approach will be explained later.

In order to calculate the contributions of two-pion exchanges, we need the following  LO and NLO $\pi N$ Lagrangians,
\begin{equation}\label{eq:eff}
\mathcal{L}=\mathcal{L}_{\pi N}^{(1)}+\mathcal{L}_{\pi N}^{(2)},
\end{equation}
where the superscript refers to the respective chiral order, and they read~\cite{Fettes:2000gb,Chen:2012nx}, respectively,
\begin{align}\label{eq:lagpiN}
\mathcal{L}_{\pi N}^{(1)} &=\bar{N}\left( {\rm{i}} \slashed{D}-m + \frac{g_A}{2} \slashed{u} \gamma_5 \right) N ,\\
\mathcal{L}_{\pi N}^{(2)} &=c_{1}\trace{\chi_{+}}\bar{N}N- \frac{c_{2}}{4m^2}\trace{u^{\mu}u^\nu}\left(\bar{N}D_{\mu}D_{\nu}N +h.c.\right)+\frac{c_3}{2}\trace{u^2}\bar{N} N-\frac{c_4}{4}\bar{N}\gamma^{\mu}\gamma^{\nu}\left[u_{\mu},u_{\nu}\right] N,
\end{align}
where the nucleon field $N=(p,n)^{T}$, and the covariant derivative {$D_\mu$} is defined as $D_{\mu}=\partial_{\mu}+\Gamma_{\mu}$ with
\begin{align}
\nonumber \Gamma_{\mu}=\frac{1}{2}\left(u^\dag \partial_{\mu} u+ u \partial_{\mu} u^\dag\right),~~~~~~~u={\rm{exp}}\left(\frac{{\rm{i}}\Phi}{2f_\pi}\right).
\end{align}
The pion field $\Phi$ is a $2 \times 2$ matrix of the following form,
\begin{align}
\nonumber\Phi=\left(
 \begin{matrix}
   \pi^0 &  \sqrt{2} \pi^{+}\\
   \sqrt{2} \pi^{-} &  -\pi^0\\
  \end{matrix}
 \right),
\end{align}
and the axial current {type quantity} $u_{\mu}$ is defined as,
\begin{align}
\nonumber u_{\mu}={\rm{i}}\left(u^\dag \partial_{\mu} u- u \partial_{\mu} u^\dag\right),
\end{align}
where $\chi_{+}=u^\dag \chi u + u \chi u^\dag$ with $\chi=\mathcal{M}=diag\left(m_\pi^2,m_\pi^2\right)$. The following values for the
  relevant LECs and masses are adopted in the numerical calculation: The pion decay constant $f_{\pi}=92.4$ MeV, the axial coupling constant $g_A=1.29$~\cite{Machleidt:2011zz}\footnote{This choice is made in order to be consistent with Refs.~\cite{Entem:2003ft,Epelbaum:2004fk,Epelbaum:2008ga,Machleidt:2011zz, Epelbaum:2014sza,Entem:2015xwa,Entem:2017gor}. Using the
   more standard value $g_A=1.267$ yields almost the same  results.}, the nucleon mass $m_n=939$ MeV, the pion mass $m_\pi=139$ MeV~\cite{Tanabashi:2018oca}, and the low-energy constants $ c_1 =-1.39$, $c_2=4.01$, $c_3=-6.61$, $c_4=3.92$, all in units of GeV$^{-1}$, taken from Ref.~\cite{Chen:2012nx} . {The values of $c_{1,2,3,4}$ employed in this work are different from the standard values (of HB ChPT) due to the different renormalization schemes adopted. In Ref.~\cite{Chen:2012nx}, the pion-nucleon scattering is studied in the EOMS scheme, which is also the scheme we adopted in this work, while previous calculations are mainly based on the HB scheme. Therefore, we took the values of $c_{1,2,3,4}$ from Ref.~\cite{Chen:2012nx} for self-consistency. We cannot use the HB values of $c_{1,2,3,4}$ because then the description of pion-nucleon scattering would be ruined.} One should note that the complete  $\mathcal{L}_{\pi N}^{(2)}$ contains more terms than what are relevant here.
\section{Two-pion exchange contributions} \label{sec:oneloop}
\subsection{ Leading order ($O(p^2)$) results }

\begin{figure}
\centering
\includegraphics[width=0.45\textwidth]{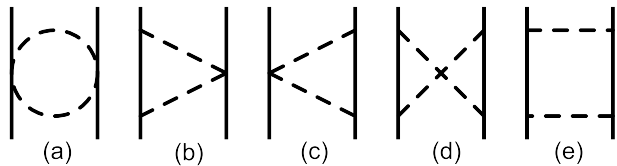}
\caption{Two-pion exchange diagrams at $O(p^2)$. The pion nucleon vertices refer to vertices from $\mathcal{L}_{\pi N}^{(1)}$}.
\label{fig:tpelo}
\end{figure}

The two-pion exchange {$T$-matrix} is evaluated in the center-of-mass frame and {in} the isospin limit  $m_u=m_d$. The leading order TPE diagrams are shown in Fig.~\ref{fig:tpelo}. They contribute to order $O(p^2)$. All of them can be calculated directly in the EOMS scheme~\cite{Gegelia:1999gf,Fuchs:2003qc}, which is just the conventional dimensional regularization scheme with further removal of PCB terms. There is no so-called pinch singularity~\cite{Weinberg:1990rz,Weinberg:1991um,Kaiser:1997mw} in this case due to the appearance of the finite nucleon mass. Note that only direct diagrams need to be computed  in the $NN$ $T$-matrix because of the Pauli exclusion principle~\cite{Kaiser:1997mw}. In principle, the box diagram includes contributions from irreducible TPE and iterated OPE.  As a matter of fact, we have calculated the iterated OPE {by inserting the relativistic OPE potential~\cite{Ren:2016jna} into the Lippmann-Schwinger equation} and found that the result is {numerically} identical to that in Ref.~\cite{Kaiser:1997mw}. For this reason, the contributions from iterated OPE are not presented explicitly.

 The complete TPE {contributions to the on-shell $NN$ $T$-matrix} are decomposed into the scalar integrals $A_0, B_0, C_0$ and $D_0$ multiplied with some polynomials and fermion bilinears using FeynCalc~\cite{Shtabovenko:2020gxv,Shtabovenko:2016sxi,Mertig:1990an} and then calculated numerically with the help of OneLOop~\cite{vanHameren:2009dr,vanHameren:2010cp}. However, there is still a tricky problem. That is, because of the large non-zero nucleon mass $m_n$ in the chiral limit,  lower order analytical terms may appear in higher order loop calculations which then break the naive power counting, namely the PCB problem~\cite{Gasser:1987rb}. The procedure to remove the PCB terms is rather standard and has been discussed in detail in Refs.~\cite{Gegelia:1999gf,Fuchs:2003qc,Geng:2013xn,Lu:2018zof}. At the end, the total TPE {contributions to the on-shell $NN$ $T$-matrix} at $O(p^2)$ take the following form,
\begin{equation}\label{eq:tpelo}
\mathcal{T}_{NN}^{(2)}=\frac{1}{16\pi^2 f_\pi^4}\sum_{i}N_i \mathcal{T}_i^{(2)},
\end{equation}
where $i$ refers to the $i$-th Feynman diagram contributing at this order, $N_i$ {denotes} the {product of isospin and coupling factors} which is summarized in Table~\ref{tb:isfoptwo}, {$\mathcal{T}_i^{(2)}$} refers to the $T$-matrix from each Feynman diagram and

\begin{table}
\centering
\caption{{Products of isospin and coupling factors} of two-pion exchange diagrams at $O(p^2)$}
\label{tb:isfoptwo}
\begin{tabular}{c|c|c|c|c|c}
\hline
\hline
 & football & triangle L & triangle R& box & crossed\\
 \hline
  $I=1$ & $\frac{1}{8 }$  & $\frac{1}{8 } g_A^2$ &$\frac{1}{8 } g_A^2$  & $\frac{1}{16  } g_A^4$  & $\frac{5}{16  }g_A^4$ \\
 \hline
 $I=0$ & $ -\frac{3}{8 }$ & $-\frac{3}{8 }g_A^2$  & $-\frac{3}{8 }g_A^2$ & $\frac{9}{16  }g_A^4$ &$ -\frac{3}{16  }g_A^4$ \\
\hline
\hline
\end{tabular}
\end{table}

\begin{equation}
\mathcal{T}_i^{(2)} = \mathcal{T}_i^{\prime(2)} - \mathcal{T}^{\prime \prime(2)}_i,
\end{equation}
where {$\mathcal{T}_i^{\prime(2)}$} denotes the total {contribution to $T$-matrix} and {$\mathcal{T}^{\prime \prime(2)}_i$} is the {contribution from the} PCB term. The total {contributions to $T$-matrix} for
the football, triangle, cross and box diagrams read explicitly
\begin{align}\label{eq:tpepotentials}
\mathcal{T}^{\prime(2)}_{\text{Football}} &= -\frac{1}{18}{\cal{B}}_2 \left[3 \left(4 m_{\pi }^2-t\right) B_0\left(t,m_{\pi }^2,m_{\pi }^2\right)+6
   \text{A}_0\left(m_{\pi }^2\right)+12 m_{\pi }^2-2 t\right],\\\nonumber
\mathcal{T}^{\prime(2)}_{\text{TrigL}} &= 4 m_n^2 \left\{{\cal{B}}_3 m_n \left[2 \left(2
   {f_1} +{f_2}+{f_3}\right) +{C}_0\left(\text{A}\right)\right]+2 {\cal{B}}_2
   {f_4}\right\} + 2{\cal{B}}_2{f_5},\\\nonumber
\mathcal{T}^{\prime(2)}_{\text{TrigR}} & = \mathcal{T}^{\prime(2)}_{\text{TrigL}}({\cal{B}}_3\mapsto {\cal{B}}_4),\\\nonumber
\mathcal{T}^{\prime(2)}_{\text{Cross}} & = -\left\{2 m_n^2 \left\{m_n \left[2\left({\cal{B}}_3+{\cal{B}}_4\right)\left({f_2} +{f_3}\right)+4 \left(4 {\cal{B}}_1 m_n+{\cal{B}}_3+{\cal{B}}_4\right) {f_1} \right.\right.\right.\\\nonumber
&\left.\left.\left.+\left(4 {\cal{B}}_1   m_n+{\cal{B}}_3\right){C}_0\left(\text{A}\right) +\left(4 {\cal{B}}_1  m_n+{\cal{B}}_4\right) {C}_0\left(\text{B}\right)\right]+4\left\{{\cal{B}}_2{f_4}+m_n^2\left[\left({\cal{B}}_3 + {\cal{B}}_4\right) \right.\right.\right.\right.\\\nonumber
  &\left.\left.\left.\left. \times   D_{22} m_n  +D_{23} \left(2 {\cal{B}}_1 m_n^2+{\cal{B}}_5\right)+2 {\cal{B}}_2 D_{00}\right]\right\}+2 {\cal{B}}_1{B_0}\left(t,m_{\pi }^2,m_{\pi}^2\right)\right\} +{\cal{B}}_2{f_5}\right\},\\\nonumber
  V^{\prime(2)}_{\text{Box}} &=-\left\{2 m_n^2 \left\{m_n \left[-2\left( {\cal{B}}_3 + {\cal{B}}_4\right)\left({f_2}+{f_3}\right) +4 \left(4 {\cal{B}}_1 m_n-{\cal{B}}_3-{\cal{B}}_4\right) {f_1} \right.\right.\right.\\\nonumber
  &\left.\left.\left.+\left(4 {\cal{B}}_1 m_n-{\cal{B}}_3\right){C}_0\left(\text{A}\right) +\left(4 {\cal{B}}_1 m_n-{\cal{B}}_4\right) {C}_0\left(\text{B}\right)\right] -4\left\{ {\cal{B}}_2{f_4} + m_n^2\left[\left({\cal{B}}_3+{\cal{B}}_4\right)\right.\right.\right.\right.\\\nonumber
  &\left.\left.\left.\left. \times D_{22} m_n  -D_{23} \left(2 {\cal{B}}_1 m_n^2+{\cal{B}}_5\right)+2 {\cal{B}}_2 D_{00}\right]\right\}+2 {\cal{B}}_1{B_0}\left(t,m_{\pi }^2,m_{\pi}^2\right)\right\}-{\cal{B}}_2{f_5}\right\},
\end{align}
with
\begin{align}
f_1 & =  {\text{PaVe}}\left(1,\left\{m_n^2,t,m_n^2\right\},\left\{m_n^2,m_{\pi}^2,m_{\pi}^2\right\}\right),\\\nonumber
f_2 & =  {\text{PaVe}}\left(1,1,\left\{m_n^2,t,m_n^2\right\}, \left\{m_n^2,m_{\pi }^2,m_{\pi}^2\right\}\right),\\\nonumber
f_3 & ={\text{PaVe}}\left(1,2,\left\{m_n^2,t,m_n^2\right\},\left\{m_n^2,m_{\pi}^2,m_{\pi}^2\right\}\right),\\\nonumber
f_4 & = {\text{PaVe}}\left(0,0,\left\{m_n^2,t,m_n^2\right\},\left\{m_n^2,m_{\pi}^2,m_{\pi}^2\right\}\right),\\\nonumber
f_5 & = {\text{PaVe}}\left(0,0,\{t\},\left\{m_{\pi }^2,m_{\pi}^2\right\}\right),\\\nonumber
C_0(\text{A}) & =  {C}_0\left(m_n^2,m_n^2,t,m_{\pi}^2,m_n^2,m_{\pi }^2\right),\\\nonumber
C_0(\text{B}) & = {C}_0\left(m_n^2,t,m_n^2,m_n^2,m_{\pi }^2,m_{\pi }^2\right),
\end{align}
where {$D_{ij}$} and  PaVe are the library functions {of} FeynCalc{~\cite{Shtabovenko:2020gxv,Shtabovenko:2016sxi,Mertig:1990an}} and can be simplified to the scalar integrals $A_0, B_0, C_0, D_0$, and ${\cal{B}}_{1-5}$ {denote} the following fermion bilinears,

\begin{align}
{\cal{B}}_1 & = \bar u(\bm{p}')u(\bm{p}) \bar u(-\bm{p}')u(-\bm{p}),\\\nonumber
{\cal{B}}_2 & = \bar u(\bm{p}') \gamma^\mu u(\bm{p}) \bar u(-\bm{p}') \gamma_\mu u(-\bm{p}),\\\nonumber
{\cal{B}}_3 & = \bar u(\bm{p}') (\slashed{p}_2+\slashed{p}_4) u(\bm{p}) \bar u(-\bm{p}')u(-\bm{p}),\\\nonumber
{\cal{B}}_4 & = \bar u(\bm{p}') u(\bm{p}) \bar u(-\bm{p}') (\slashed{p}_1+\slashed{p}_3) u(-\bm{p}),\\\nonumber
{\cal{B}}_5 & = \bar u(\bm{p}') (\slashed{p}_2 +\slashed{p}_4 )u(\bm{p}) \bar u(-\bm{p}') \slashed{p}_1 u(-\bm{p}),\\\nonumber
\end{align}
where $\bm{p}, \bm{p}' $are incoming and outgoing three momentum, $p_1^\mu=(E, \bm p)$, $p_2^\mu=(E,-\bm {p})$, $p_3^\mu=(E',\bm {p}')$, $p_4^\mu=(E',-\bm{ p}' )$, $E=\sqrt{\bm{p}^2+m_n^2}$, $E'=\sqrt{\bm{p}'^2+m_n^2}$, $t=(p_1-p_3)^2$, $u(\bm {p})$ and $\bar{u}(\bm {p})$ are Dirac spinors,
\begin{align}
u(\bm {p}) = N \left(
\begin{matrix} \mathbbm{1}\\
\frac{{\bm{\sigma}} \cdot \bm{p}}{E+m_n}
\end{matrix}
\right) \chi_s,~~~~~~~N = \sqrt{\frac{E+m_n}{2m_n}},
\end{align}
where $\chi_s$ {denotes} the {Pauli} spinor matrix and ${\bm{\sigma}}$ is the {Pauli} matrix. To obtain the {$T$-matrix from the} PCB terms, it is  convenient to project the {$T$-matrix} from momentum space to helicity space so that the {$T$-matrix} become scalar without the {Pauli} matrix and can easily be expanded in powers of small parameters. The detailed procedure to do this projection is explained later. However, because of their complexity, we do not show the explicit expressions of {$\mathcal{T}'$} in helicity space here.   One thing to be noted is that {the above bilinears are all parity even.} This can be easily shown by utilizing momentum conservation and the Dirac equation, e.g.,
\begin{align}
\nonumber {\cal{B}}_5 &= \bar u(\bm{p}') \left(\slashed{p}_2 + \slashed{p}_4 \right)u(\bm{p}) \bar u(-\bm{p}') \slashed{p}_1 u(-\bm{p}) \\\nonumber
&= \bar u(\bm{p}') \left(\slashed{p}_2 + \slashed{p}_4\right) u(\bm{p})\bar u(-\bm{p}') \frac{1}{2}\left(\slashed{p}_1 + \slashed{p}_3+\slashed{p}_4 - \slashed{p}_2\right) u(-\bm{p}) \\\nonumber
&= \frac{1}{2}\bar u(\bm{p}')\left(\slashed{p}_2 +  \slashed{p}_4\right) u(\bm{p}) \bar u(-\bm{p}') \left(\slashed{p}_1 + \slashed{p}_3\right) u(-\bm{p})\\\nonumber
&\xrightarrow{\text{Parity}} {\cal{B}}_5.
\end{align}

The {$T$-matrices from the} PCB terms in helicity space read,

\begin{align}\label{eq:pcblo}
\mathcal{T}^{\prime \prime(2)}_{\text{Football}} & = 0,\\\nonumber
\mathcal{T}^{\prime \prime(2)}_{\text{TrigL}} & = 4{H_1} m_n^2 \ln \left(\frac{\mu}{m_n}\right),\\\nonumber
\mathcal{T}^{\prime \prime(2)}_{\text{TrigR}} & = \mathcal{T}_{\text{TrigL}}^{\prime \prime},\\\nonumber
\mathcal{T}^{\prime \prime(2)}_{\text{Cross}} & = -4H_1 m_n^2 \left[3 \ln \left(\frac{\mu}{m_n}\right)-1\right],\\\nonumber
\mathcal{T}^{\prime \prime(2)}_{\text{Box}}   & =- 4H_1 m_n^2 \left[ \ln \left(\frac{\mu}{m_n}\right)+1\right] ,\\\nonumber
\end{align}

where {$\mu$} refers to the renormalization scale { and is set to 1 GeV  in our numerical study unless otherwise stated}, $H_{1}$ is,
\begin{align}
H_1 = \left[|\bar{\lambda}_1 +\lambda_1| \cos\left(\frac{\theta}{2}\right) +|\bar{\lambda}_1 -\lambda_1| \sin\left(\frac{\theta}{2}\right)  \right] \left[ |\bar{\lambda}_2 +\lambda_2| \cos\left(\frac{\theta}{2}\right) - |\bar{\lambda}_2 -\lambda_2| \sin\left(\frac{\theta}{2}\right) \right],
\end{align}
where $\lambda_{1,2},\bar{\lambda}_{1,2}$ denote the helicities of incoming, outgoing particles respectively and $\theta$ refers to the scattering angle.

\subsection{Next-to-leading order ($O(p^3)$) results}

\begin{figure}
\centering
\includegraphics[width=0.45\textwidth]{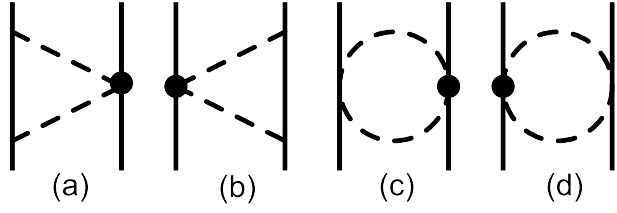}
\caption{Two-pion exchange diagrams at $O(p^3)$. The black dots denote vertices from $\mathcal{L}_{\pi N}^{(2)}$}.
\label{fig:tpenlo}
\end{figure}

The next-to-leading order TPE diagrams are shown in Fig.~\ref{fig:tpenlo}. These diagrams are the same as the corresponding
diagrams shown in Fig.~\ref{fig:tpelo} with the replacement of the $\pi N$ vertices $\mathcal{L}_{\pi N}^{(1)}$ with $\mathcal{L}_{\pi N}^{(2)}$. Note that there is no box diagram or cross diagram at this order because there is no $\pi NN$ vertex at order $O(p^2)$.

 The next-to-leading order TPE {contributions to $T$-matrix} read,
 \begin{equation}
 \mathcal{T}_{NN}^{(3)} = \mathcal{T}_{\text{FootballL}}^{(3)}+ \mathcal{T}_{\text{FootballR}}^{(3)}+ \mathcal{T}_{\text{TrigL}}^{(3)}+ \mathcal{T}_{\text{TrigR}}^{(3)},
 \end{equation}
 where the notation is the same as that stated above. The {products of isospin and coupling factors} have been included in the {$T$-matrix}  at this order for simplicity.
 \begin{align}\label{eq:nlotpe}
 \mathcal{T}_{\text{FootballL}}^{(3)\prime} &= \frac{(4I-3)c_4}{72} \left(2 {\cal{B}}_2 m_n-{\cal{B}}_4\right) \left[3 \left(t-4 m_{\pi }^2\right)
   B_0\left(t,m_{\pi }^2,m_{\pi }^2\right)-6 {A}_0\left(m_{\pi }^2\right)+2 \left(t-6
   m_{\pi }^2\right)\right],\\\nonumber
\mathcal{T}_{\text{FootballR}}^{(3)\prime} &= \mathcal{T}_{\text{FootballL}}^{(3)\prime}({\cal{B}}_4\rightarrow {\cal{B}}_3) ,\\\nonumber
\mathcal{T}_{\text{TrigL}}^{(3)\prime} &= 6 c_1 m_n m_\pi^2 {\cal{B}}_1 \left[ 2m_n^2\left(2 f_1 + C_0(\text{A})  \right) + B_0(t,m_\pi^2,m_\pi^2)  \right]+...\quad,\\\nonumber
\mathcal{T}_{\text{TrigR}}^{(3)\prime} &= 6 c_1 m_n m_\pi^2 {\cal{B}}_1 \left[ 2m_n^2\left(2 f_1 + C_0(\text{A})  \right) + B_0(t,m_\pi^2,m_\pi^2)  \right]+...\quad .
 \end{align}
{The $T$-matrices from PCB terms in helicity space read,}
 \begin{align}
  \mathcal{T}_{\text{FootballL}}^{(3)\prime\prime} &= 0,\\\nonumber
  \mathcal{T}_{\text{FootballR}}^{(3)\prime\prime} &= 0,\\\nonumber
  \mathcal{T}_{\text{TrigL}}^{(3)\prime\prime} &= 6 c_1 H_1 m_n m_\pi^2 \left[ 2\ln \left( \frac{\mu}{m_n}\right) + 1\right]+ ...\quad,\\\nonumber
  \mathcal{T}_{\text{TrigR}}^{(3)\prime\prime} &=  6c_1 H_1 m_n m_\pi^2 \left[ 2\ln \left( \frac{\mu}{m_n}\right) + 1\right]+ ... \quad.
 \end{align}
where $I=0,1$ refers to the isospin. The full expressions of the {$T$-matrix from} the triangle {diagrams} at this order are not given explicitly due to their complexity~\footnote{They could be obtained as a Mathematica notebook from the
authors upon request.}.

 In order to compute the phase shifts, we  need to transform the {$T$-matrix} into $LSJ$ basis where $L$ is the total orbital angular momentum, $S$ is the total spin, and $J$ is the total angular momentum. The procedure for {the partial wave projection} is rather standard~\cite{Erkelenz:1971caz,Erkelenz:1974uj}. Here we refer to Ref.~\cite{Erkelenz:1974uj} for more details. At first, we compute the {$T$-matrix} directly in momentum space. Then, we transfer it to helicity basis. Next, it is rotated to the total angular momentum space $|JM\rangle$ using the Wigner d-functions. Last, it is projected to $LSJ$ basis. In order to compute the phase shifts and mixing angle, we follow the procedure of Ref.~\cite{Gasser:1990ku},
 \begin{align}
 \delta_{LSJ} &=-\frac{m_n^2 |\bm{p}|}{16\pi^2 E}\text{ Re}\langle LSJ| \mathcal{T}_{NN} |LSJ\rangle,\\\nonumber
 \epsilon_J &= \frac{m_n^2 |\bm{p}|}{16\pi^2 E}\text{ Re}\langle J-1,1,J| \mathcal{T}_{NN} |J+1,1,J\rangle.
 \end{align}
Note that the kinematical prefactors here differ from those of Ref.\cite{Kaiser:1997mw} {due
to a different sign convention for the $NN$ $T$-matrix and a different
normalization of the matrix elements obtained by partial wave
projection.}
\section{Results and discussions} \label{sec:results}
In this section, the phase shifts with $2\leq L\leq 6$ and mixing angles with $2\leq J\leq 6$ in the relativistic framework are presented and compared with those of the nonrelativistic results of Ref.~\cite{Kaiser:1997mw}. {But before showing the phase shifts, the choice of the LECs $c_{1,2,3,4}$ needs to be clarified. As we stated above, the values of $c_{1,2,3,4}$ adopted in this work are larger than those of Ref.~\cite{Bernard:1996gq} because of different renomalization schemes. In Ref.~\cite{Chen:2012nx}, the four LECs are determined by fitting to $\pi N$ scattering phase shifts in the EOMS approach at order $O(p^3)$, while in Ref.~\cite{Bernard:1996gq}, they are fixed from  a fit to nine $\pi N$ observables in the HB scheme at one-loop order $O(p^3)$.  For the sake of self-consistency, we thereby took the values of $c_{1,2,3,4}$ from Ref.~\cite{Chen:2012nx} for the EOMS case and those
of Ref.~\cite{Bernard:1996gq} for the HB case. Otherwise the descriptions of $\pi N$ scattering in both schemes will be ruined. As a matter of fact, we have performed the nonrelativistic calculation with $c_{1,2,3,4}$ fixed at the values of Ref.~\cite{Chen:2012nx} and found that the corresponding results are worse than those of Ref.~\cite{Kaiser:1997mw}. }
\subsection{D-wave}
\begin{figure}[htbp]
\centering
\subfloat{
\includegraphics[width=0.45\textwidth]{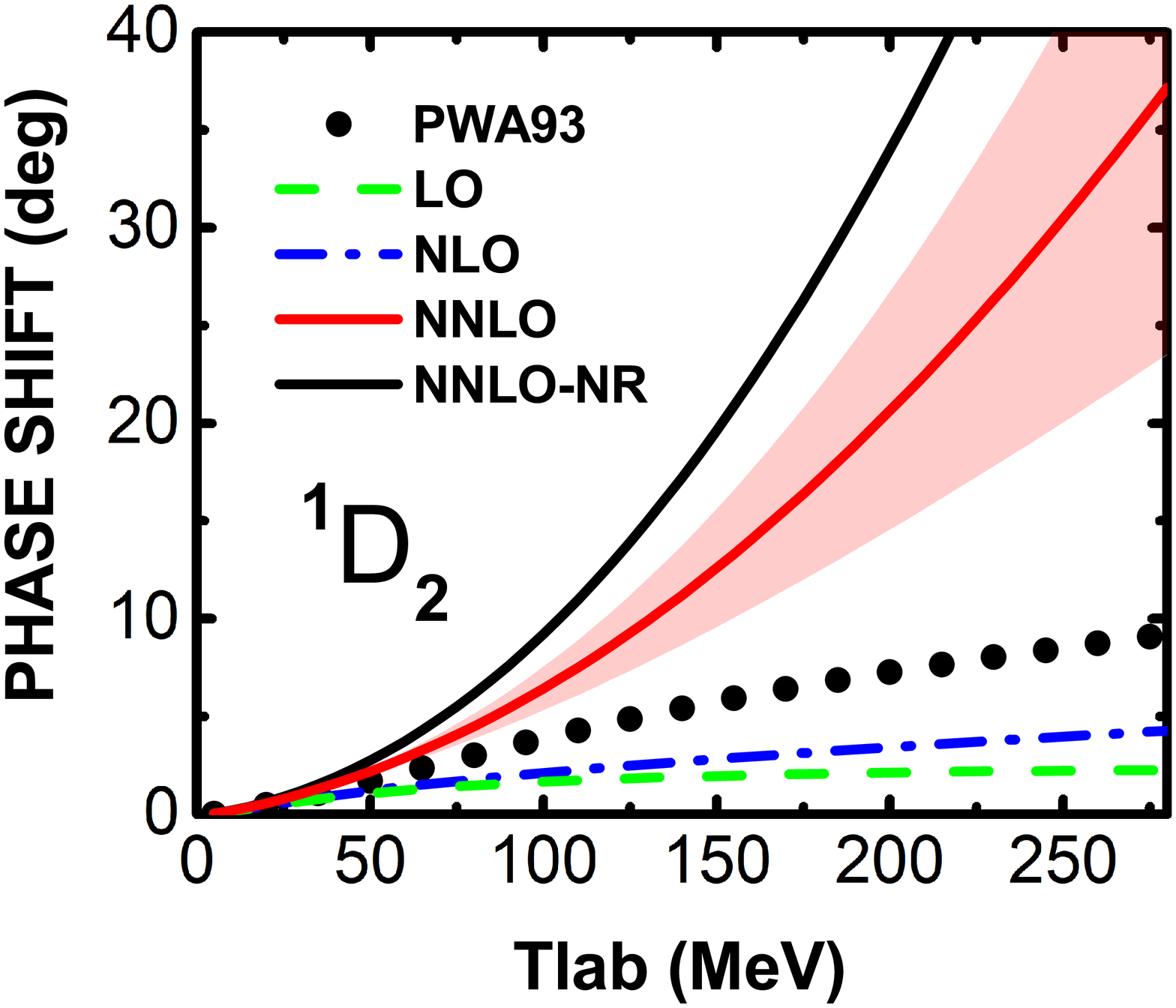}
}
\quad
\subfloat{
\includegraphics[width=0.45\textwidth]{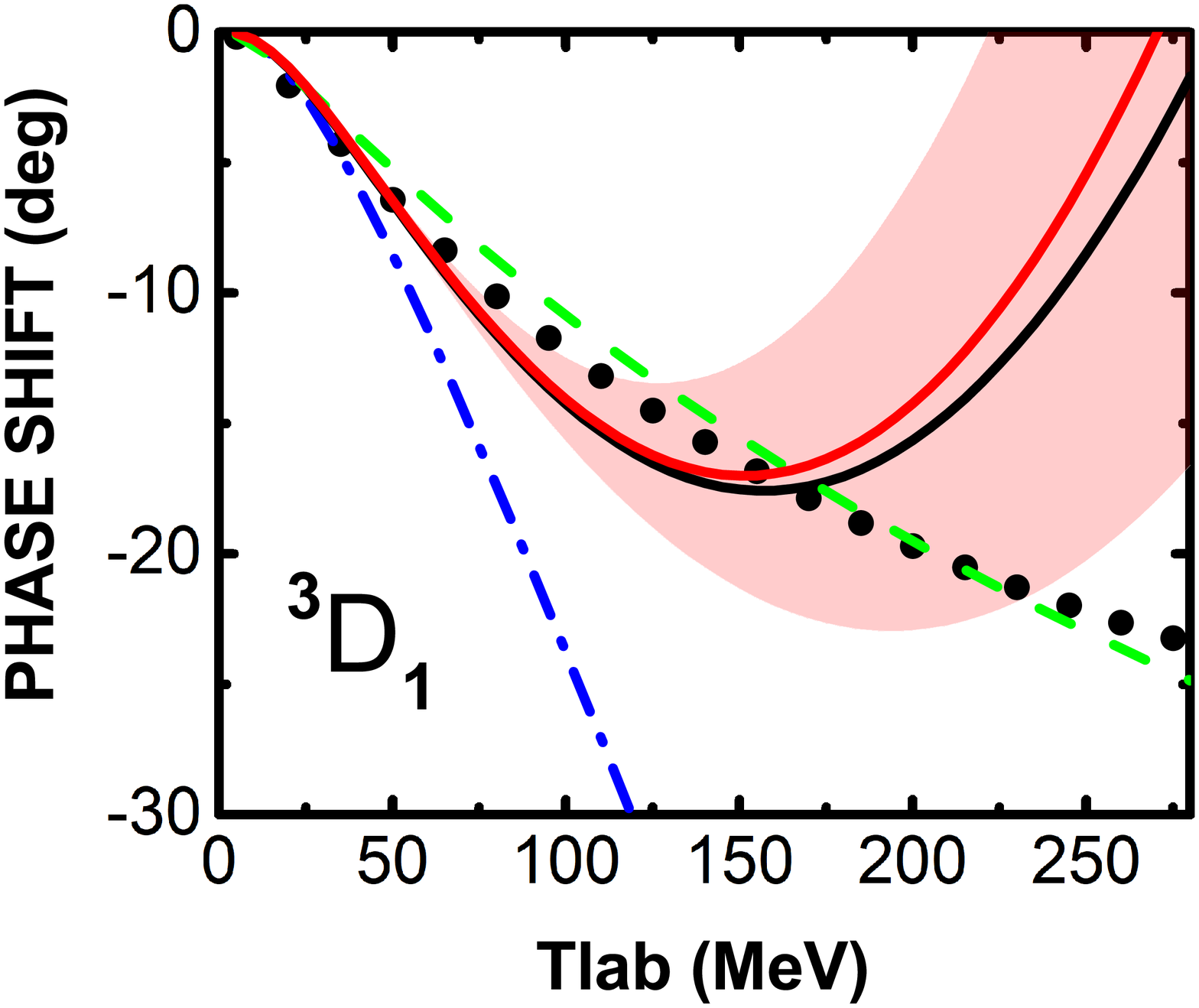}
}
\quad
\subfloat{
\includegraphics[width=0.45\textwidth]{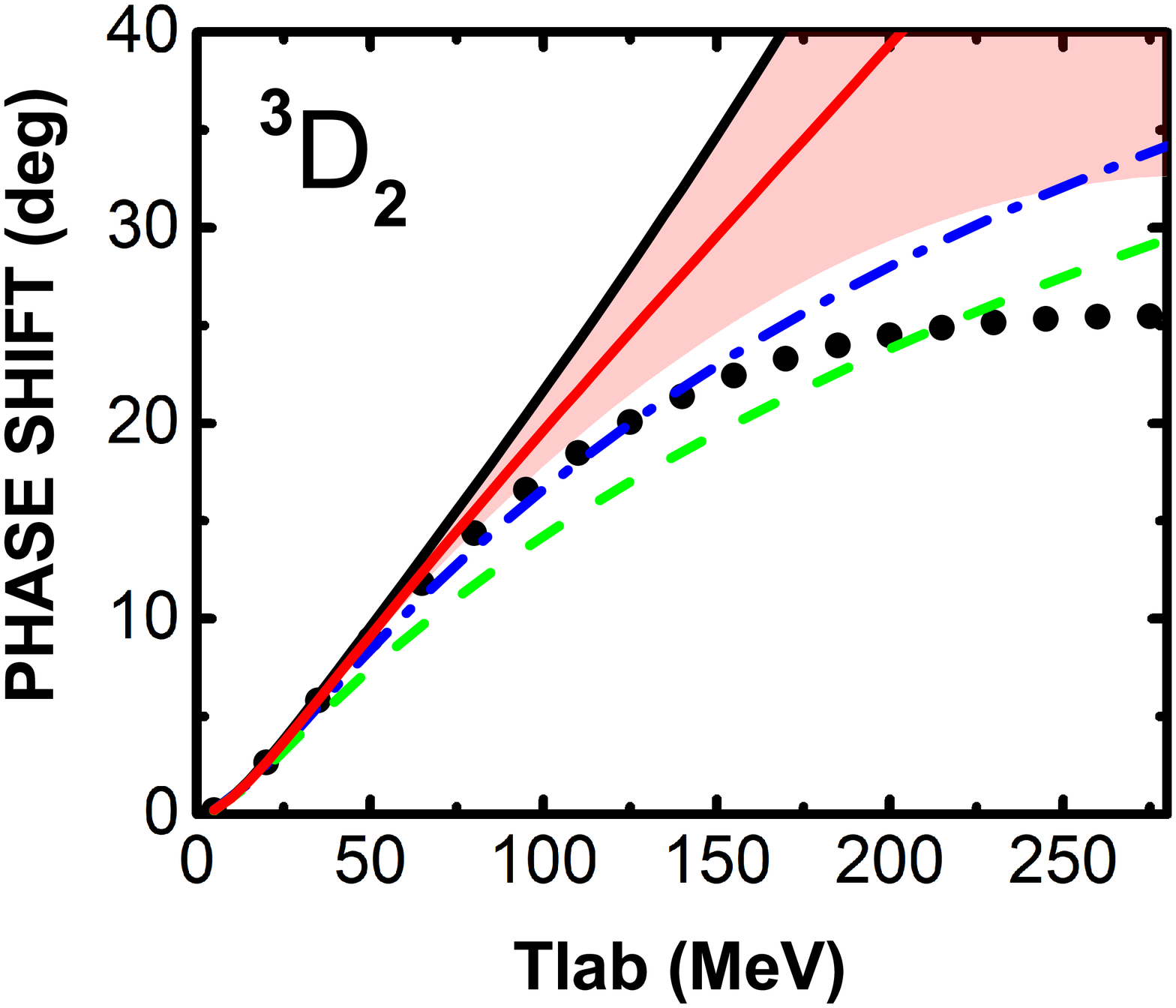}
}
\quad
\subfloat{
\includegraphics[width=0.45\textwidth]{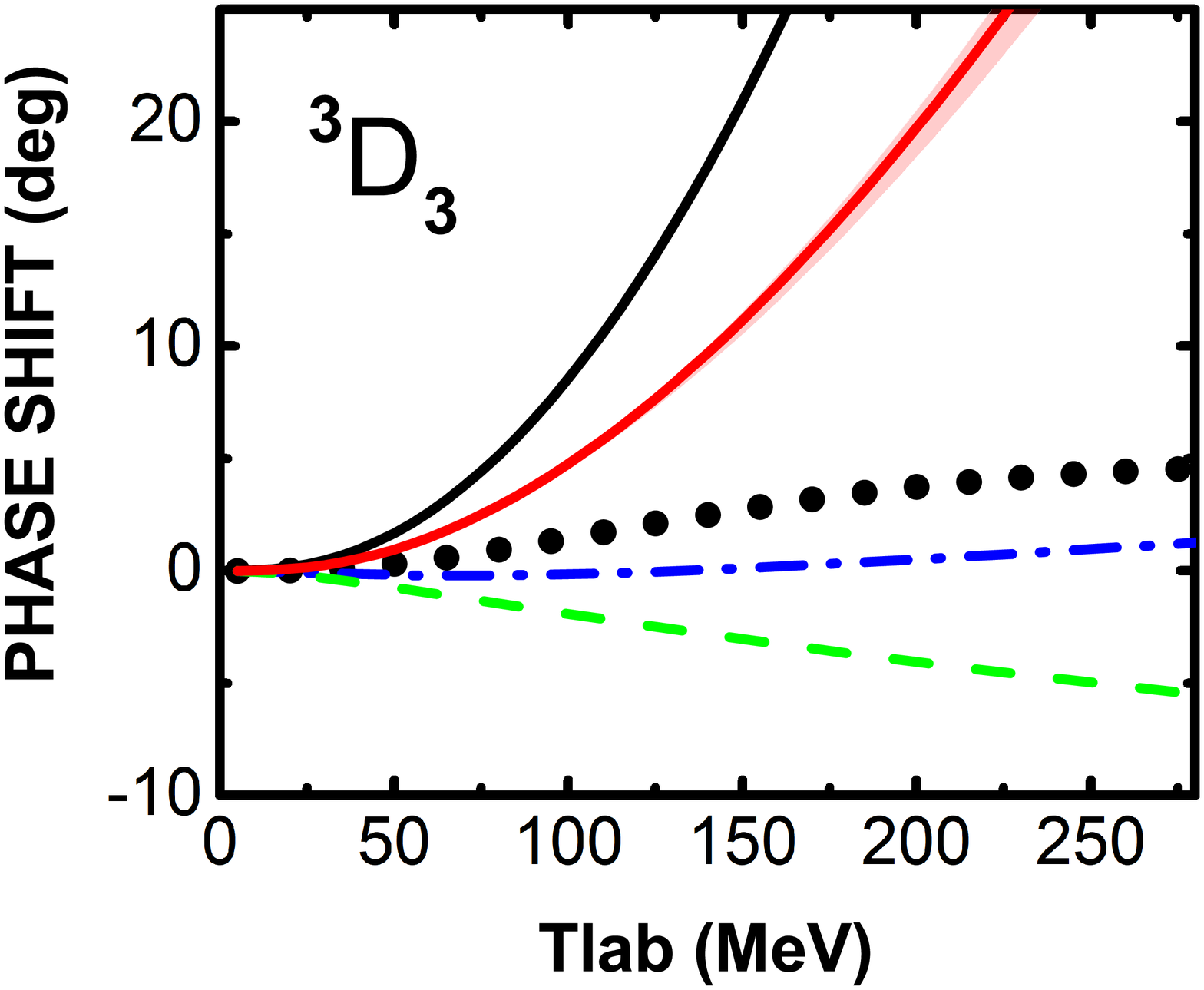}
}
\quad
\subfloat{
\includegraphics[width=0.45\textwidth]{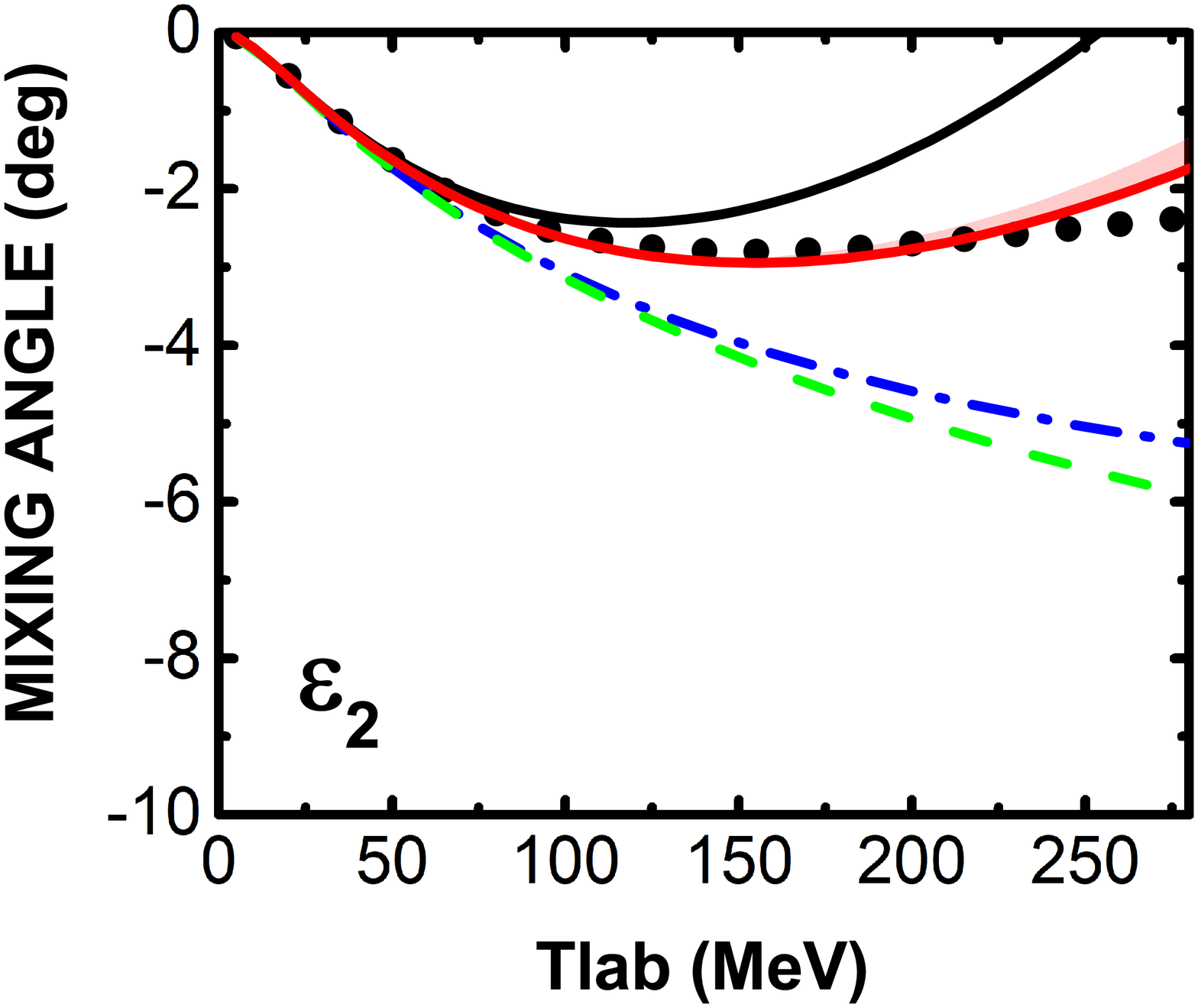}
}
\caption{D-wave phase shifts and mixing angle $\epsilon_2$ as a function of $T_{\text{lab}}$. The black dots refer to the Nijmegen partial wave phase shifts~\cite{Stoks:1993tb}. The green dashed curves correspond to the contributions from relativistic OPE~\cite{Ren:2016jna}, the blue dash dotted curves represent the contributions of leading order TPE, the red solid curves contain the next-to-leading order TPE while the black curves are their nonrelativistic counterparts~\cite{Kaiser:1997mw} with $g_A$ fixed at 1.29. {The bands are generated by varying $\mu$ from 0.5 GeV to 1.5 GeV.}}
\label{tb:Dwave}
\end{figure}

The D-wave phase shifts and mixing angle $\epsilon_2$ are shown in Fig.~\ref{tb:Dwave}. {The black dots refer to the Nijmegen partial wave phase shifts. The Green dashed lines refer to the contributions from relativistic OPE, the blue dash dotted lines represent the contributions from the leading order TPE, the red curves contain the next-to-leading order TPE, while the black curves are their nonrelativistic counterparts with $g_A$ fixed at 1.29. The bands are generated by varying $\mu$ from 0.5 GeV to 1.5 GeV. The relativistic OPE is independent of $\mu$ since it contributes at tree level. The relativistic leading order TPE is also independent of $\mu$ for partial waves with $L \geq 2$. Nonetheless, the next-to-leading order relativistic TPE depends considerably on $\mu$ for the $^3\text{D}_1$, $^1\text{D}_2$ and $^3 \text{D}_2$ partial waves and  shows little  dependence on $\mu$ for the $^3\text{D}_3$ partial wave and mixing angle $\epsilon_2$. On the other hand, the nonrelativistic results are independent of $\mu$~\cite{Kaiser:1997mw}.} For all the cases the chiral $NN$ phase shifts are in good agreement with data up to $T_{\text{lab}}=50$ MeV, and the relativistic results show the same tendency as their nonrelativistic counterparts, but the TPE contributions are more moderate, so for all the D-waves the relativistic results are in better agreement with the Nijmegen phase shifts perhaps with the exception of $^{3}\text{D}_1$ for {$T_{\text{lab}} \geq 150$ MeV}, where both descriptions show a much stronger u-turn shape, inconsistent with data.  The nonrelativistic result for $^{3}\text{D}_1$ is in fair agreement with data up to $T_{\text{lab}}=200$ MeV due to the cancellation of irreducible TPE and iterated OPE ~\cite{Kaiser:1997mw}, while in the relativistic case, the contribution of {the next-to-leading order TPE is somehow a little bit larger than the nonrelativistic counterparts so that the curve shifts somewhat upwards.} The mixing angle $\epsilon_2$ in the relativistic method is in better agreement with data due to {the moderate contribution from the irreducible parts of TPE.} Although the relativistic corrections are sizeable in D-wave and improve the description of data, the still relatively large discrepancy indicates the need of  short-range contributions, namely the contact terms controlled by LECs.

A few words are in order for the convergence pattern. For the coupled channels, because of the cancellation of the irreducible part and the iterated part in the leading order TPE, the contribution of the next-to-leading order TPE is very large compared with that of the leading order TPE. But for the singlet channel $^1\text{D}_2$, contrast to our expectation, the iterated part contributes negligibly to the phase shifts while the next-to-leading order TPE contributes a lot. Moreover, the contribution of the next-to-leading order TPE seems to be larger than the contribution of OPE. All in all, although the relativistic results are quantitatively better than the nonrelativistic results, pion-exchange contributions alone are not enough to explain the D-wave data, as concluded in Ref.~\cite{Kaiser:1997mw}.

\subsection{F-wave}
\begin{figure}[htbp]
\centering
\subfloat{
\includegraphics[width=0.45\textwidth]{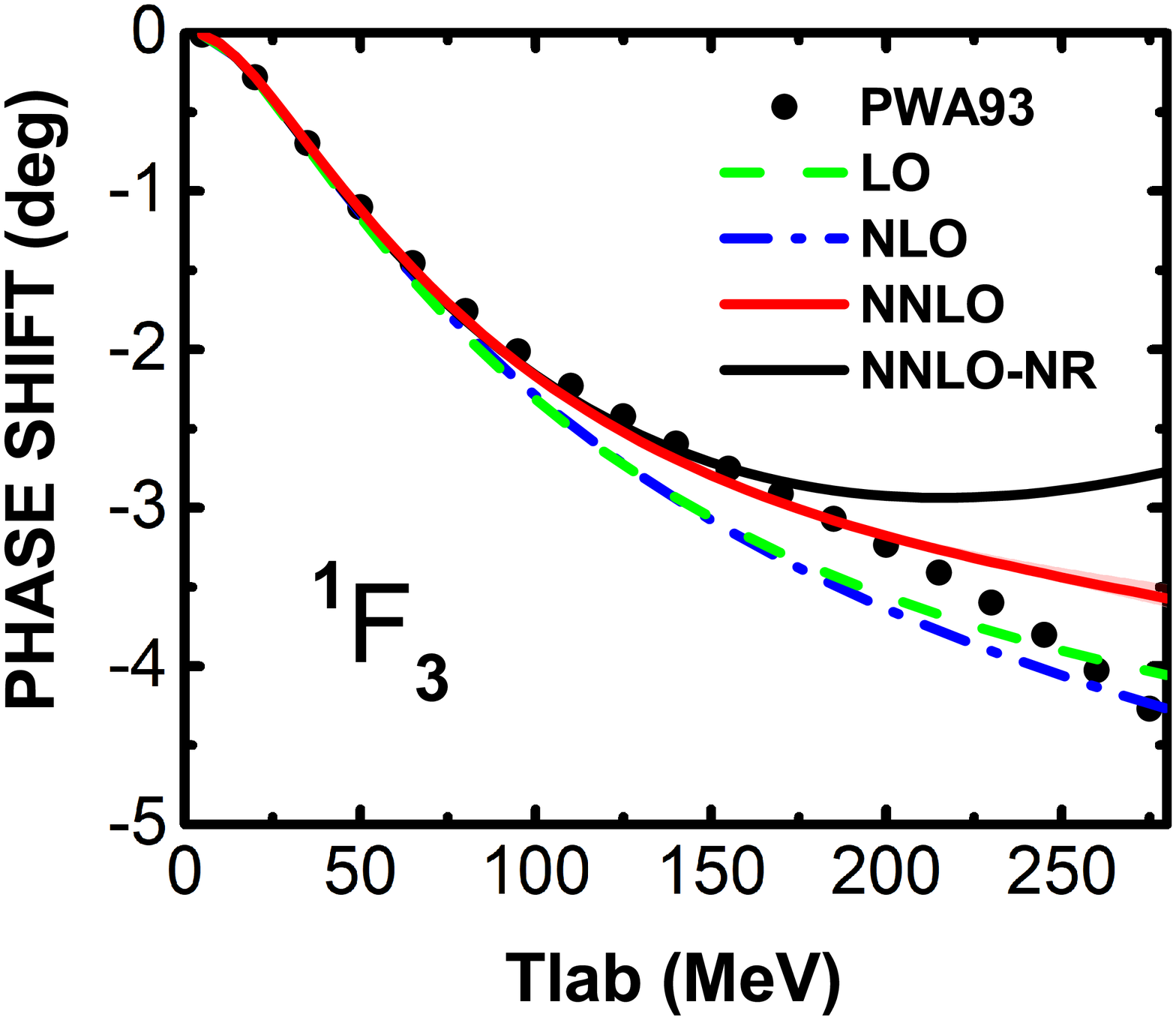}
}
\quad
\subfloat{
\includegraphics[width=0.45\textwidth]{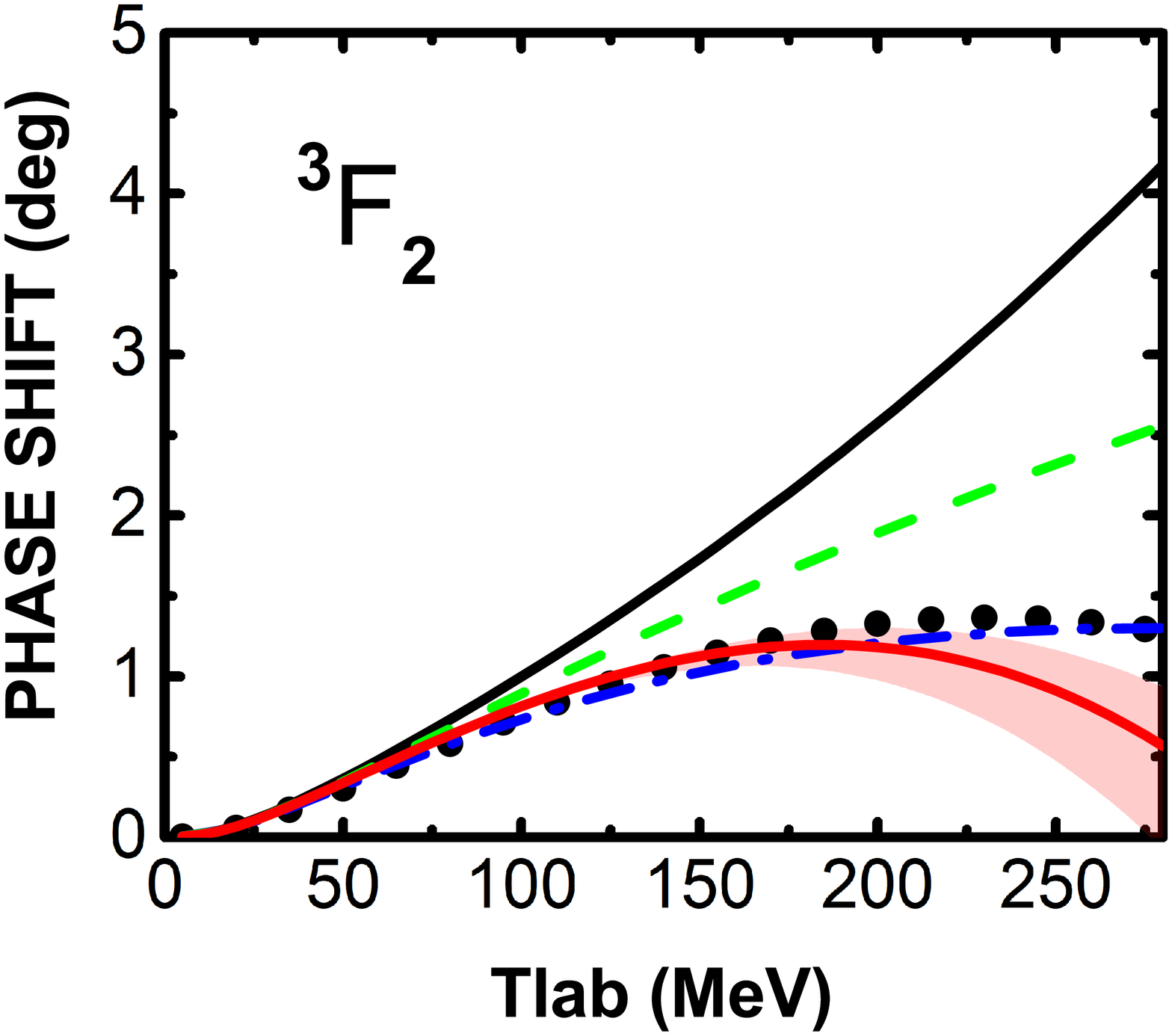}
}
\quad
\subfloat{
\includegraphics[width=0.45\textwidth]{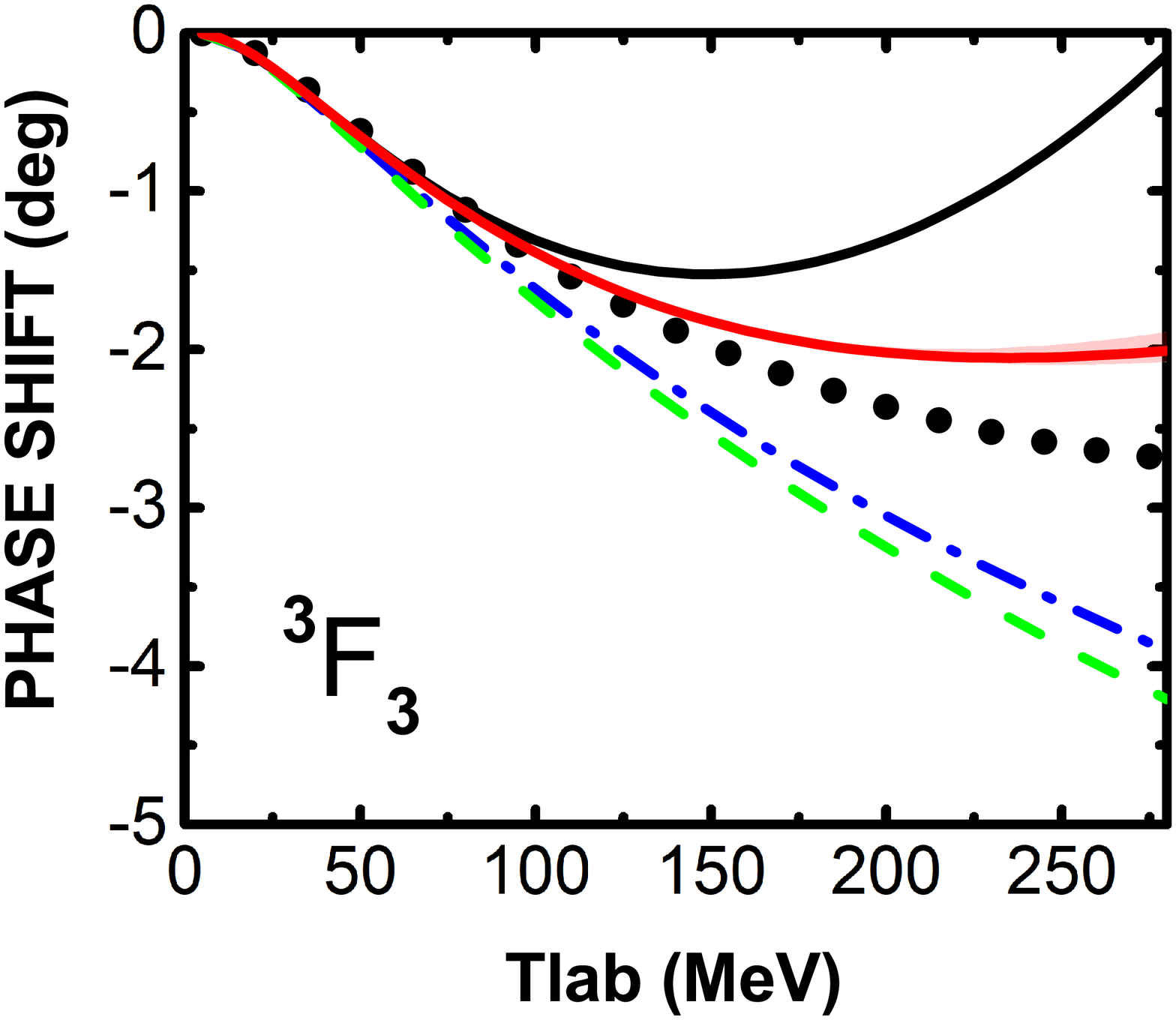}
}
\quad
\subfloat{
\includegraphics[width=0.45\textwidth]{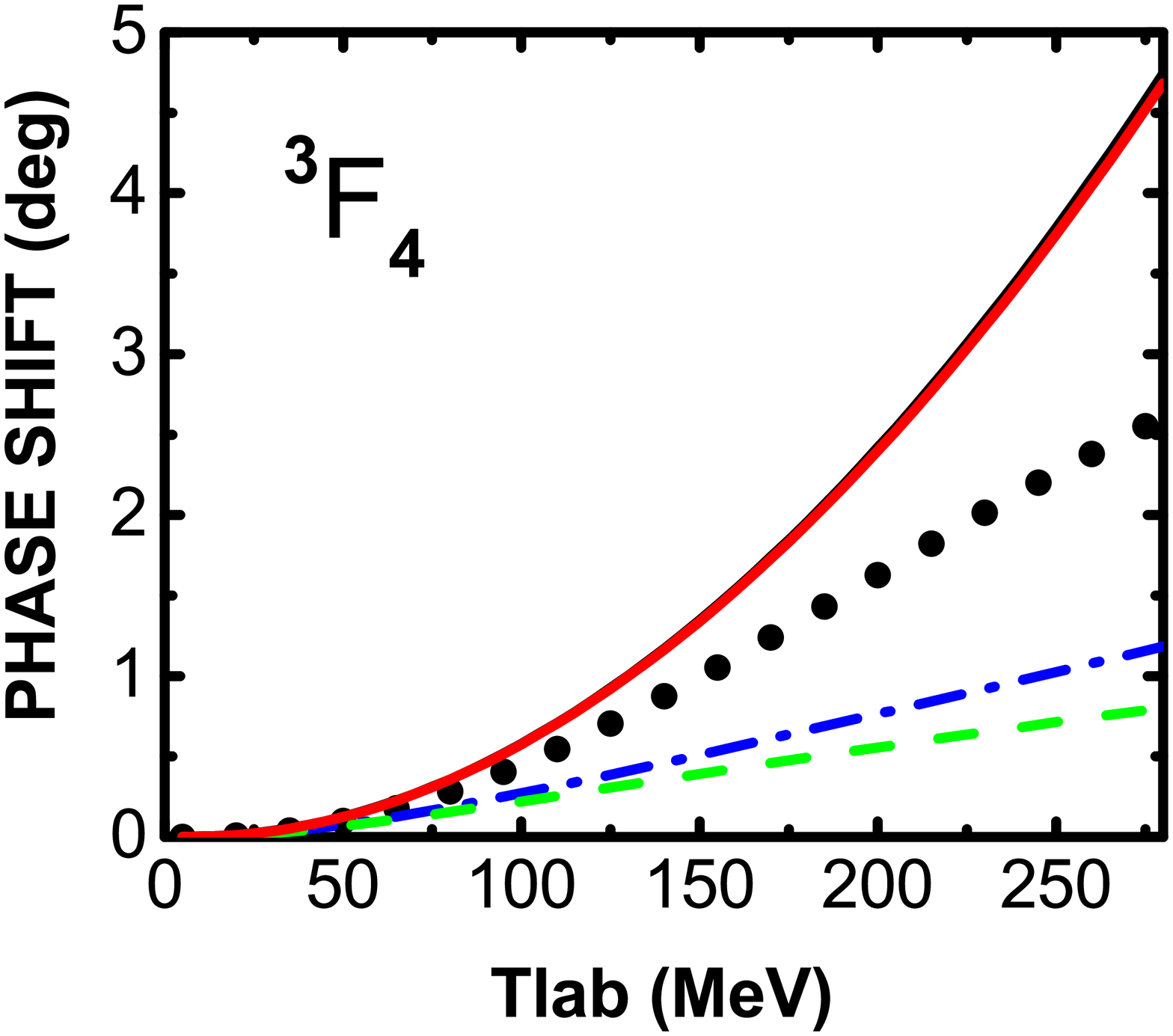}
}
\quad
\subfloat{
\includegraphics[width=0.45\textwidth]{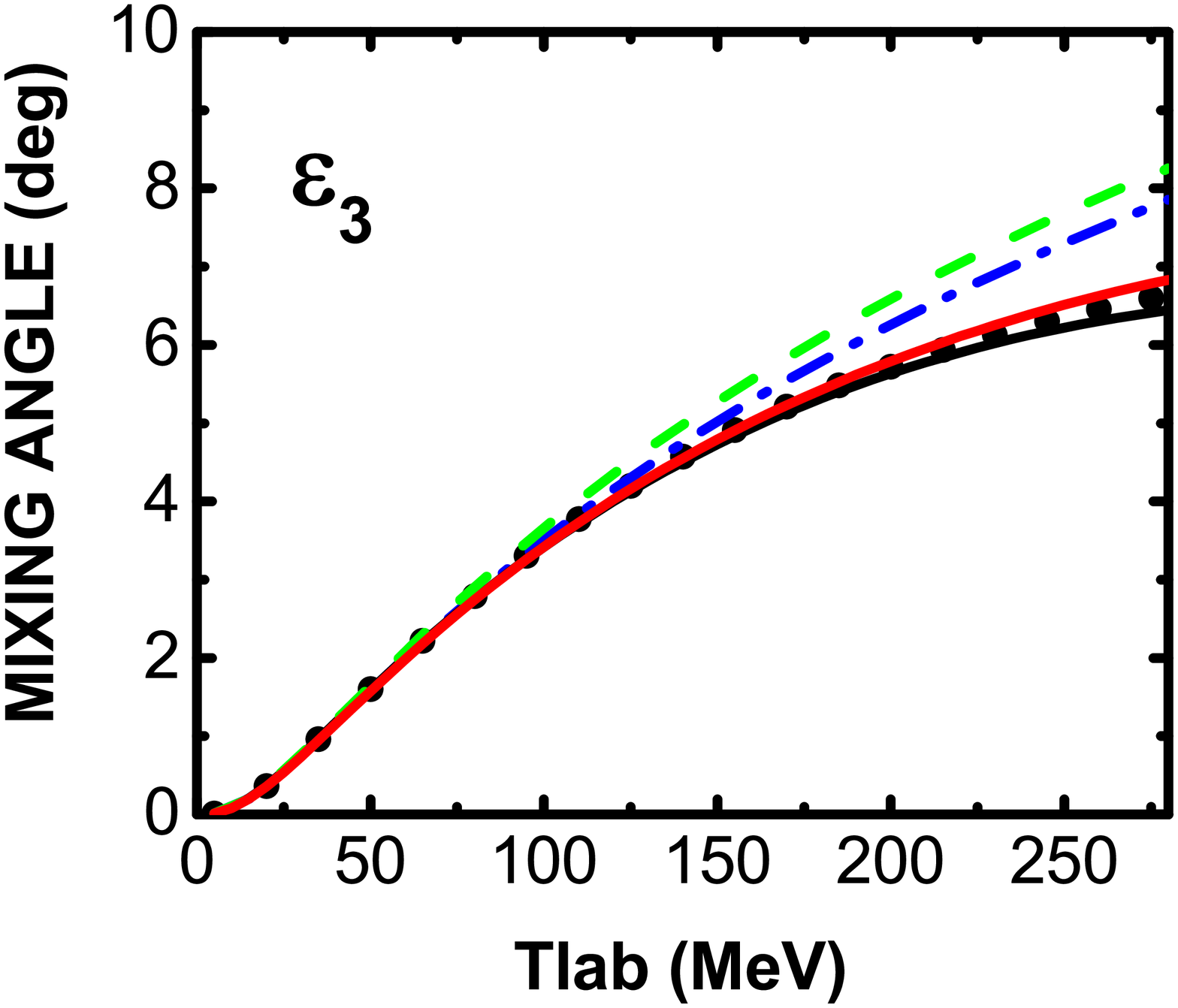}
}
\caption{Same as Fig.~\ref{tb:Dwave}, but for F-wave phase shifts and mixing angle $\epsilon_3$.}
\label{tb:Fwave}
\end{figure}

The F-wave phase shifts and mixing angle $\epsilon_3$ are depicted in Fig.~\ref{tb:Fwave}. {The relativistic chiral phase shifts only show moderate dependence on $\mu$ for the $^3\text{F}_2$ partial wave and the variation of $\mu$ yields indistinguishable difference for the others.} As in the D-wave case, the relativistic TPE is moderate so that overall the phase shifts are in better agreement with data. For the $^1\text{F}_3$ partial wave, the relativistic results are almost identical to data up to $T_{\text{lab}}=200$ MeV. For the $^3\text{F}_3$ partial wave, the relativistic phase shifts are slightly better than the nonrelativistic ones. For the $^3\text{F}_4$ partial wave, the two results are almost identical. However, for the $^3\text{F}_2$ partial wave, {the contributions from the next-to-leading order TPE are very small due to the  cancellation of the contributions from $c_3$ and $c_4$} which leads to a fair agreement with data for {$T_{\text{lab}}\leq210 $ MeV}.  In addition, the fact that the contributions of leading order relativistic TPE are relatively small indicates a good convergence at least up to $O(p^2)$. The contributions of the next-to-leading order TPE are a bit large for the {$^1\text{F}_3$}, $^3\text{F}_3$ and $^3\text{F}_4$ partial waves when $T_{\text{lab}}\geq150$ MeV {because of the large contribution from $c_3$, while the contributions of the next-to-leading order TPE are still larger than the leading order case for the mixing angle $\epsilon_3$ due to the relatively large contribution from $c_4$}. This may not be too surprising because for this energy, the momentum transfer $q$ is already about $3.85m_\pi \approx 530 $ MeV and therefore may not be regarded as a good low energy scale.

\subsection{G-wave}
\begin{figure}[htbp]
\centering
\subfloat{
\includegraphics[width=0.45\textwidth]{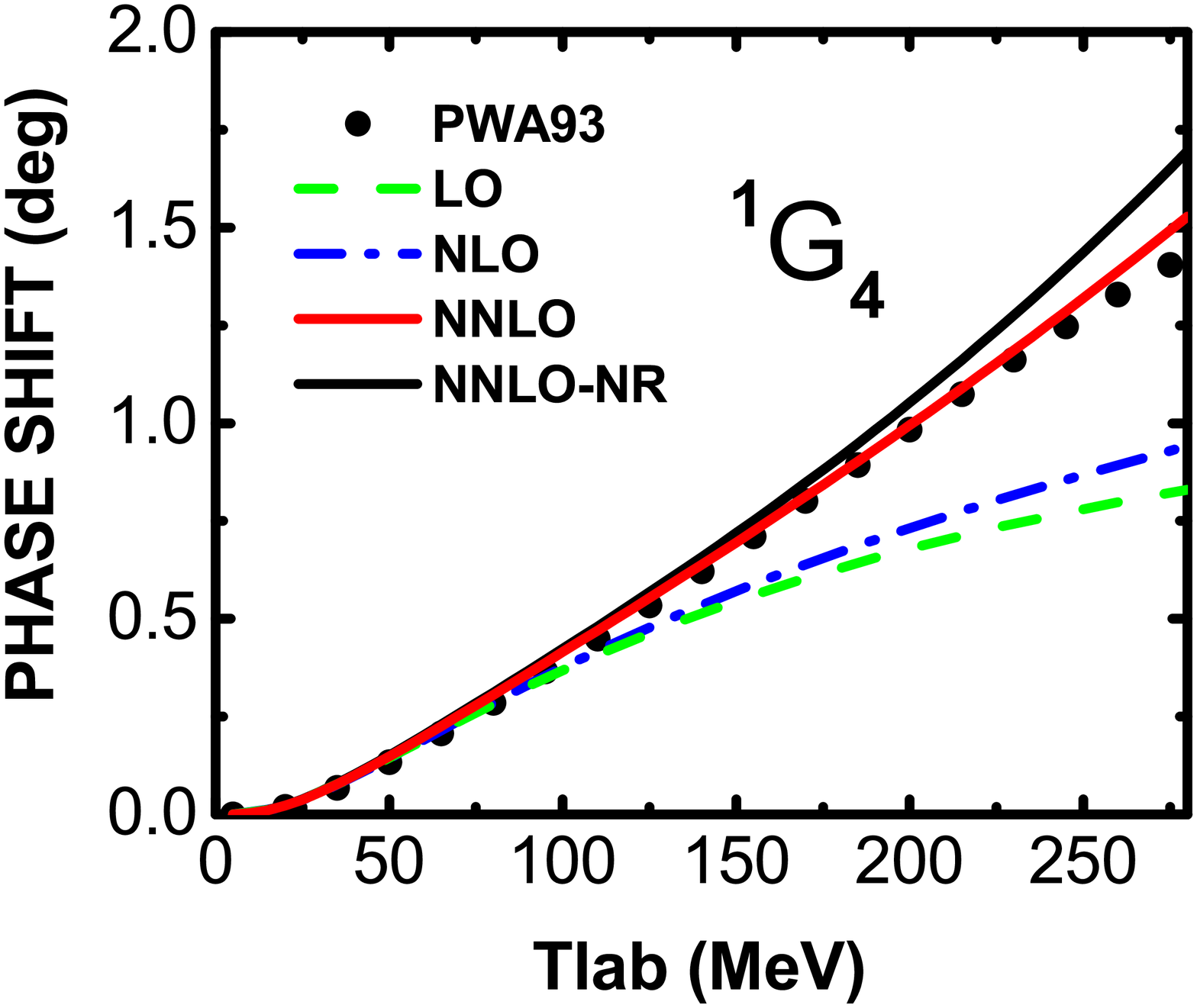}
}
\quad
\subfloat{
\includegraphics[width=0.45\textwidth]{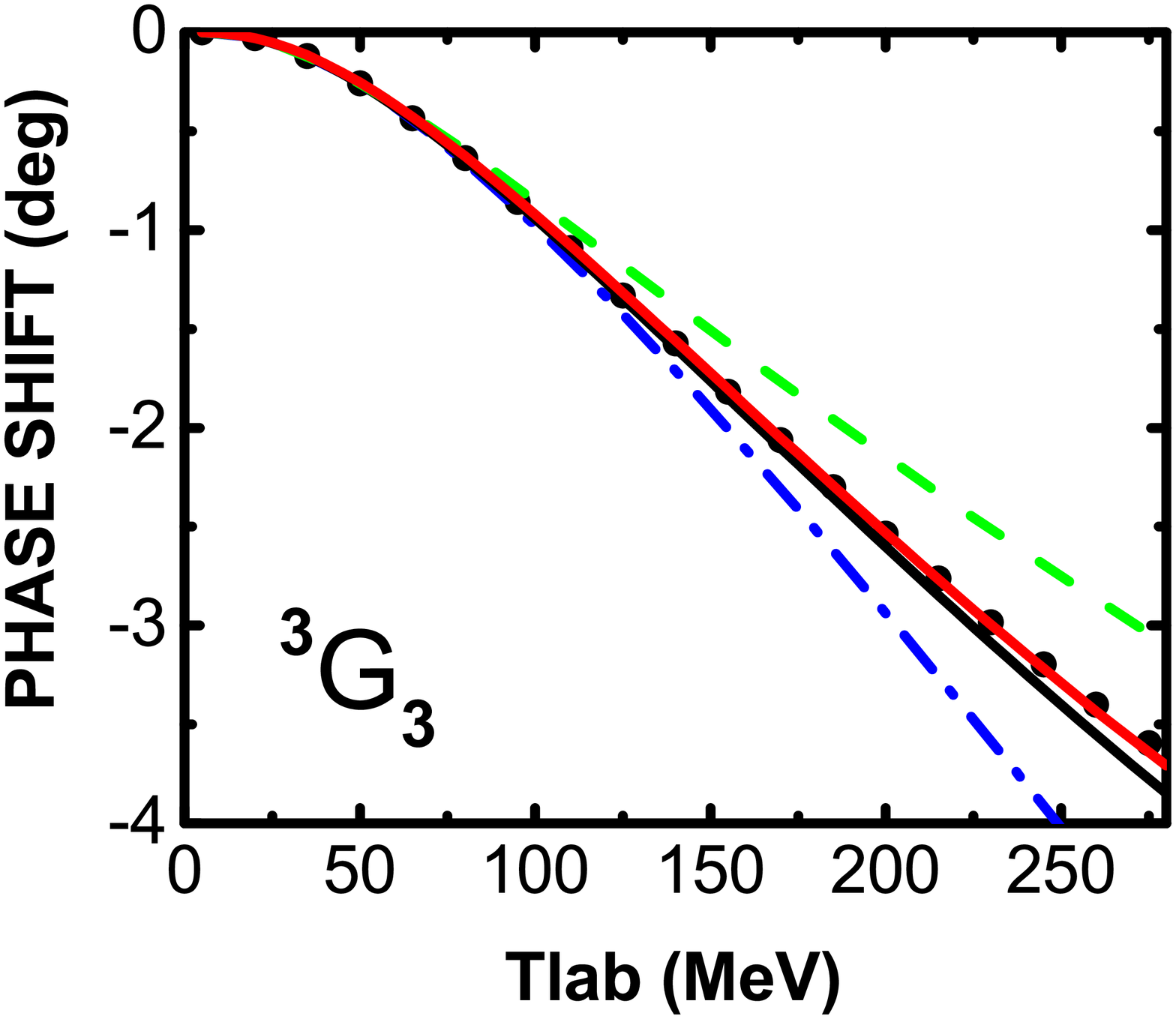}
}
\quad
\subfloat{
\includegraphics[width=0.45\textwidth]{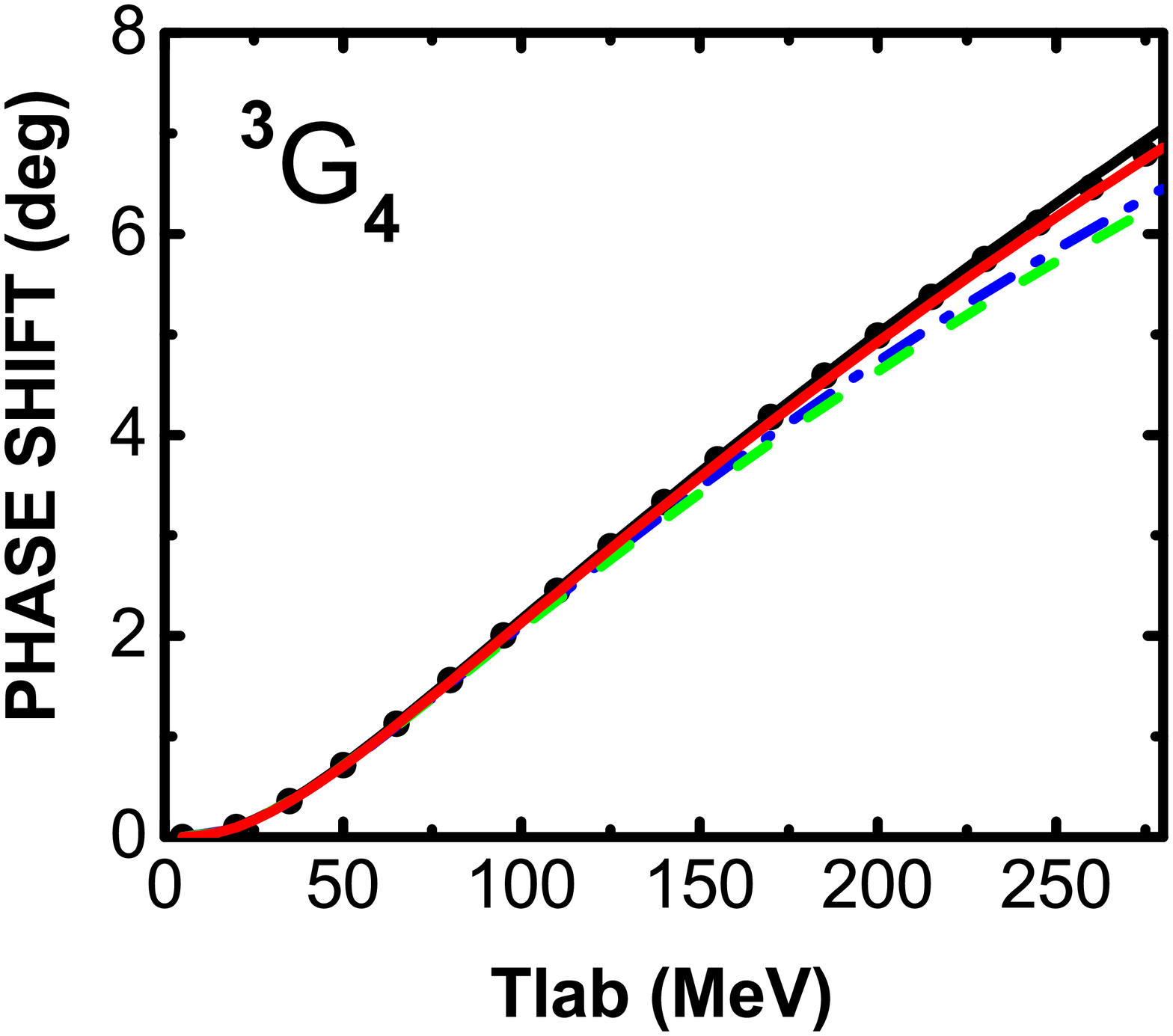}
}
\quad
\subfloat{
\includegraphics[width=0.45\textwidth]{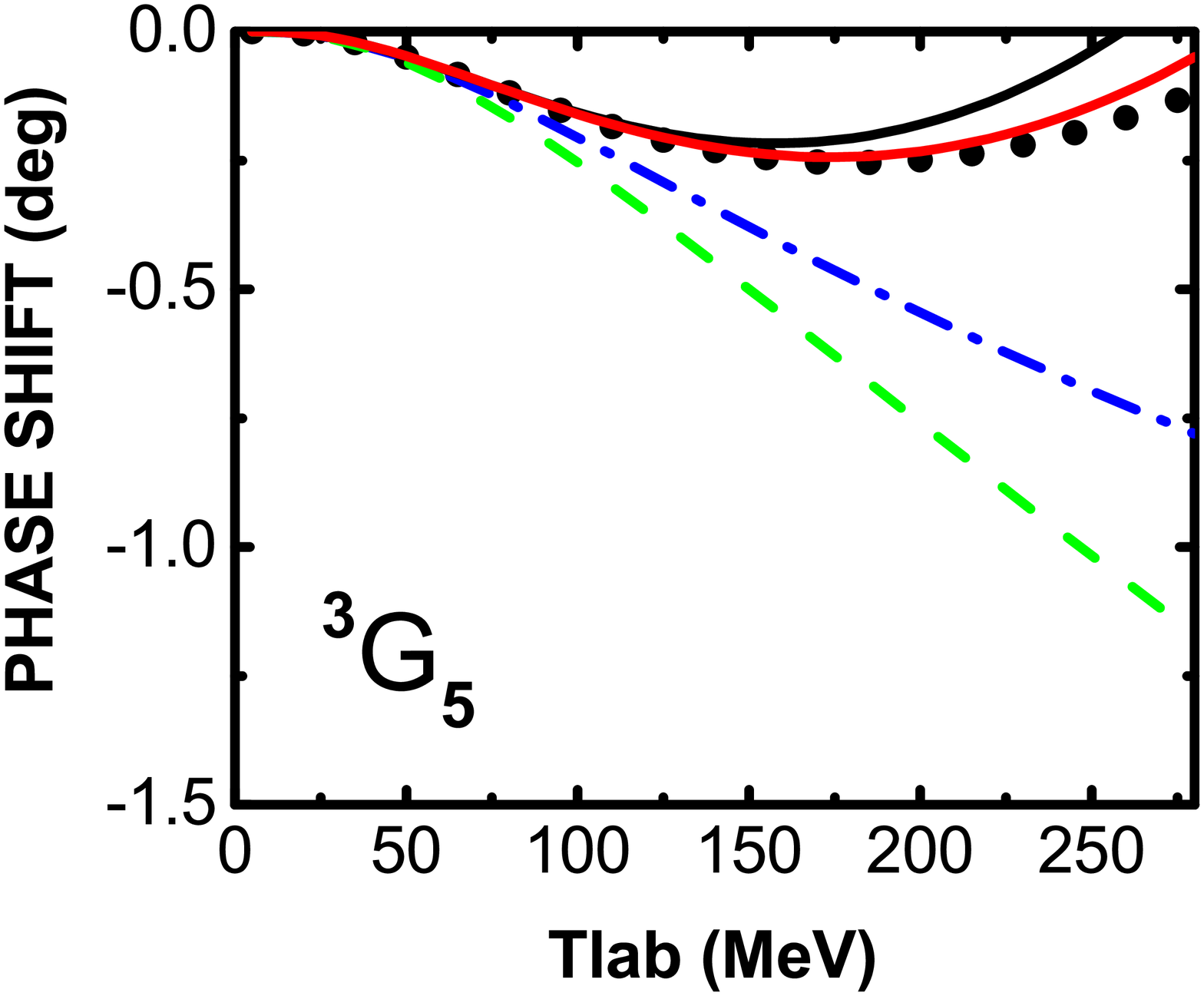}
}
\quad
\subfloat{
\includegraphics[width=0.45\textwidth]{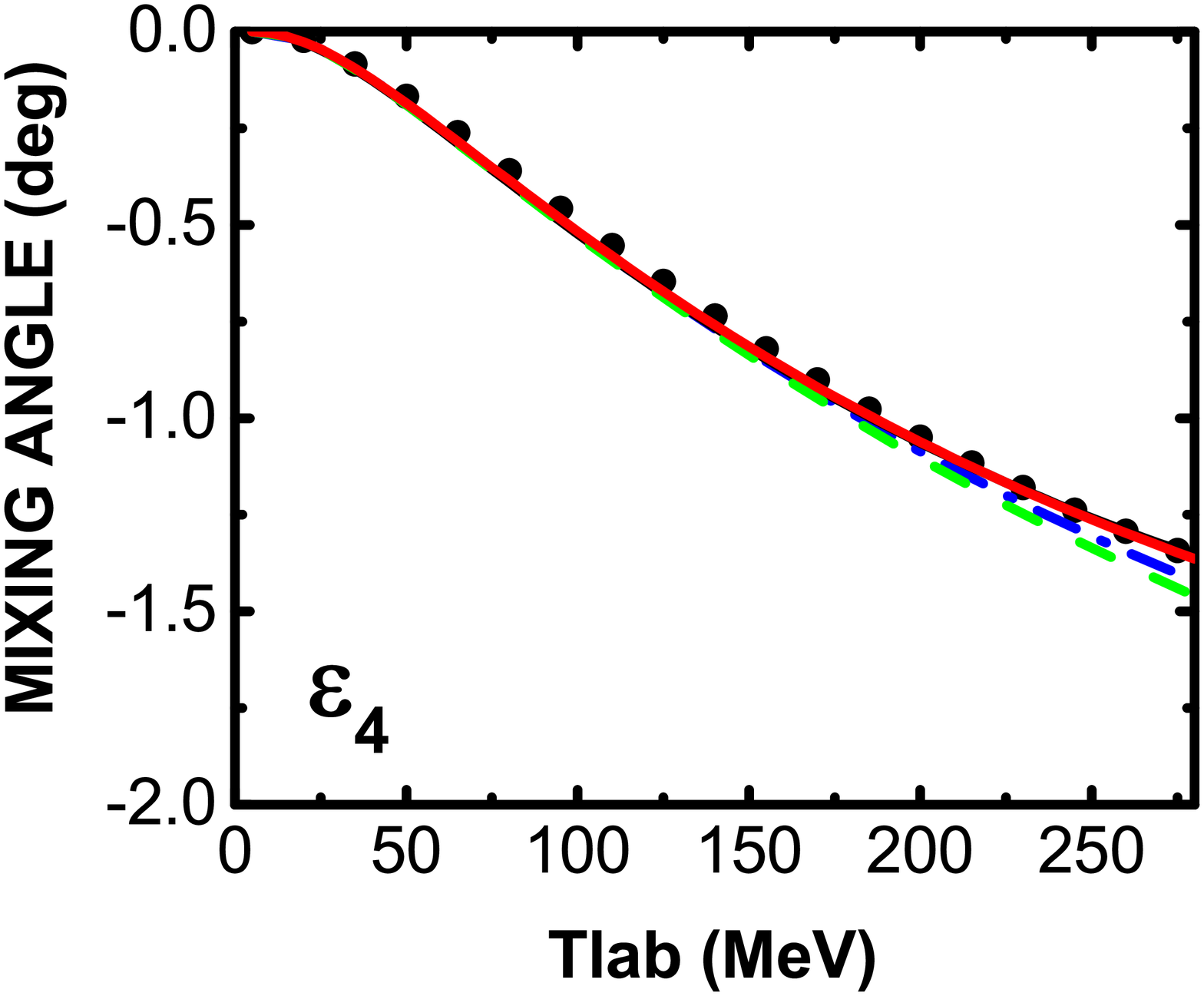}
}
\caption{Same as Fig.~\ref{tb:Dwave}, but for G-wave phase shifts and mixing angle $\epsilon_4$.}
\label{tb:Gwave}
\end{figure}

The G-wave phase shifts and mixing angle $\epsilon_4$ are depicted in Fig.~\ref{tb:Gwave}. Again, {the variation of $\mu$ affects little the phase shifts for all the partial waves and} for all the cases the relativistic phase shifts are in better agreement with data. For the $^1\text{G}_4$, $^3\text{G}_4$, {$^3\text{G}_3$} partial waves and mixing angle $\epsilon_4$, the relativistic results are almost identical to data up to $T_{\text{lab}}=280$ MeV. For the {$^3\text{G}_5$} partial waves, the relativistic phase shift is also in perfect agreement with data up to  {$T_{\text{lab}}=250$ MeV}. {Although the descriptions of phase shifts are much improved with the inclusion of the next-to-leading order TPE, the contributions from the next-to-leading order TPE are still relatively large compared with the leading order TPE for the $^1\text{G}_4$, $^3\text{G}_4$, $^3\text{G}_5$ partial waves and mixing angle $\epsilon_4$ because of the same reasons as explained in the F wave case.}

\subsection{H-wave}
\begin{figure}[htbp]
\centering
\subfloat{
\includegraphics[width=0.45\textwidth]{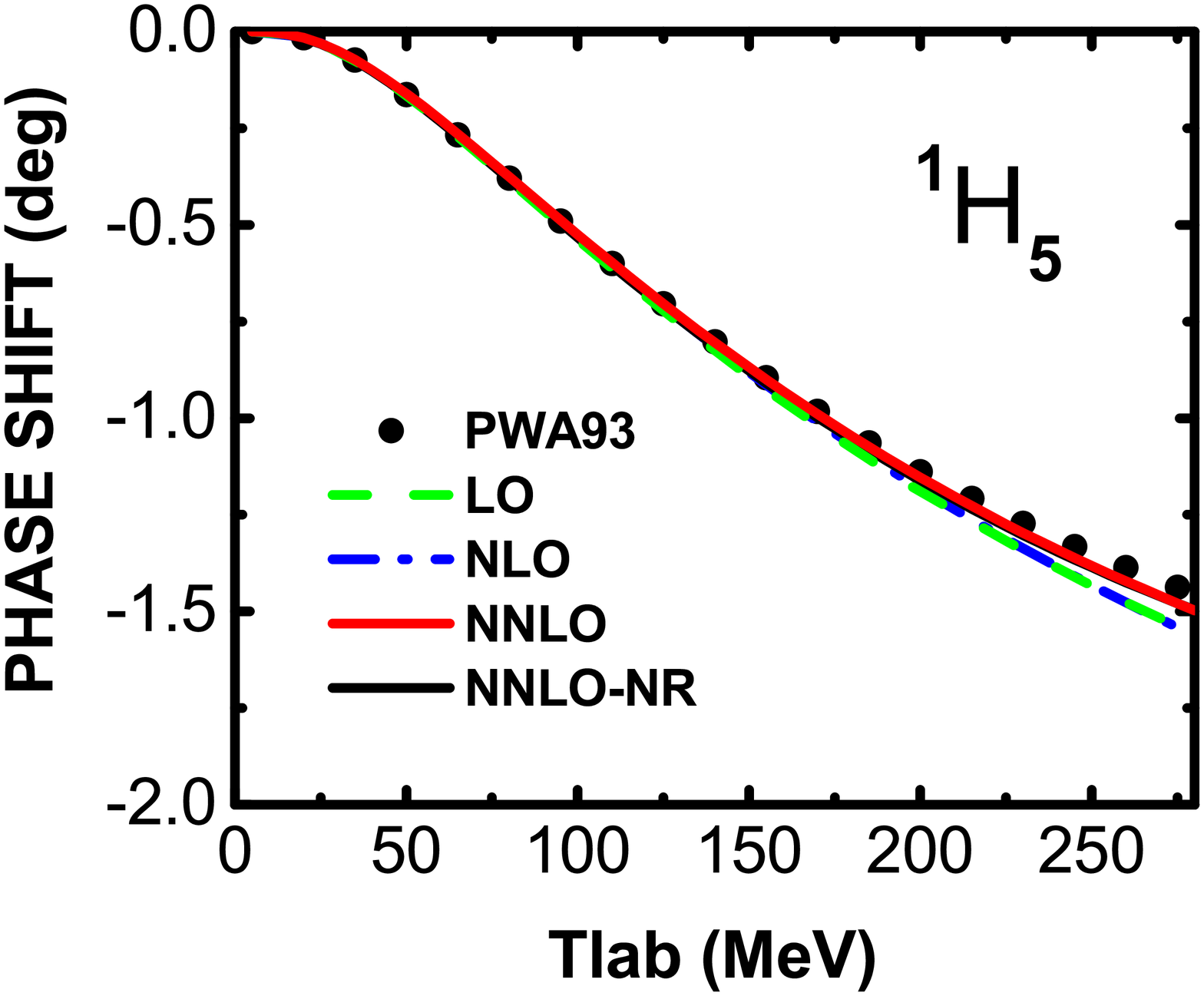}
}
\quad
\subfloat{
\includegraphics[width=0.45\textwidth]{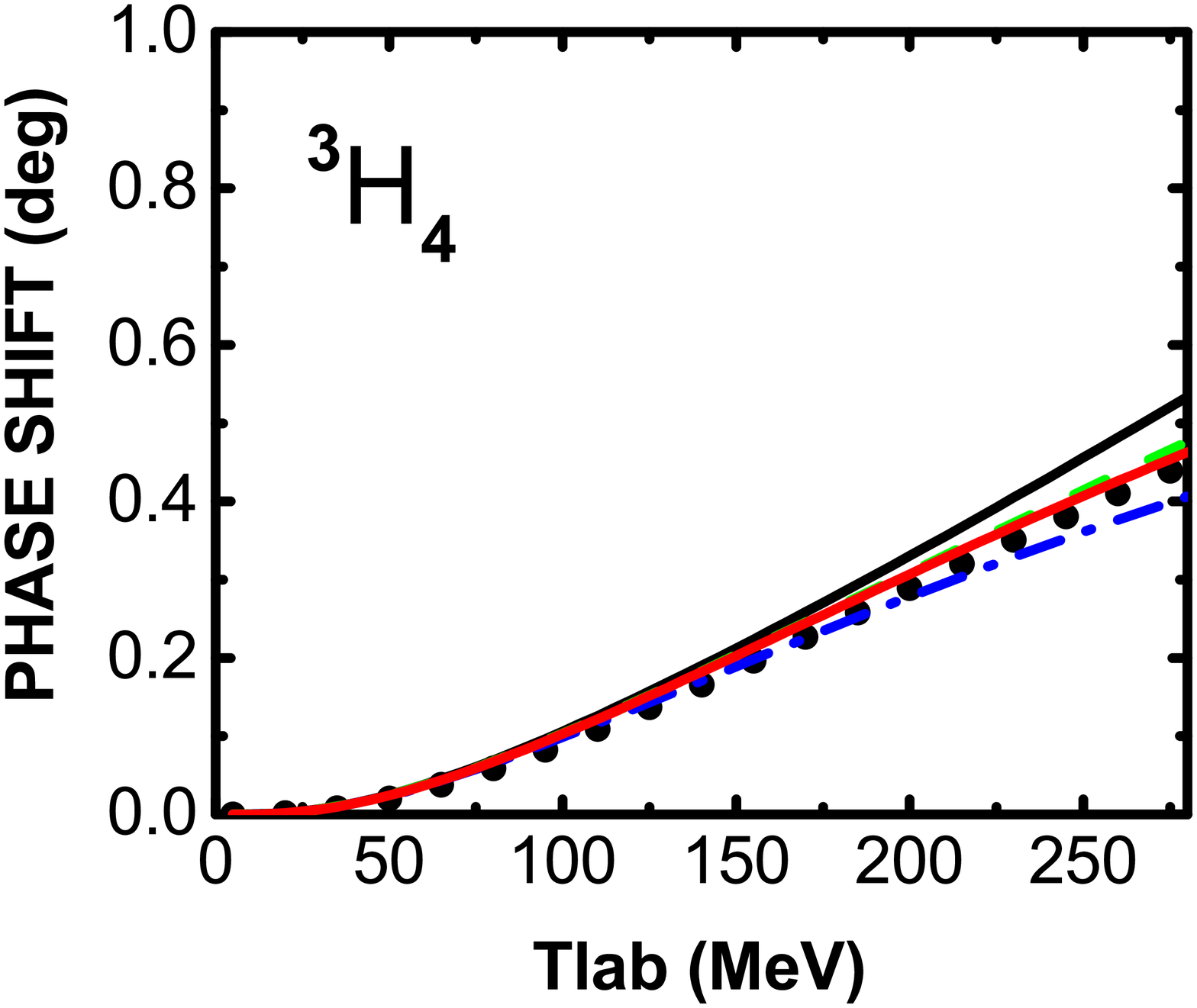}
}
\quad
\subfloat{
\includegraphics[width=0.45\textwidth]{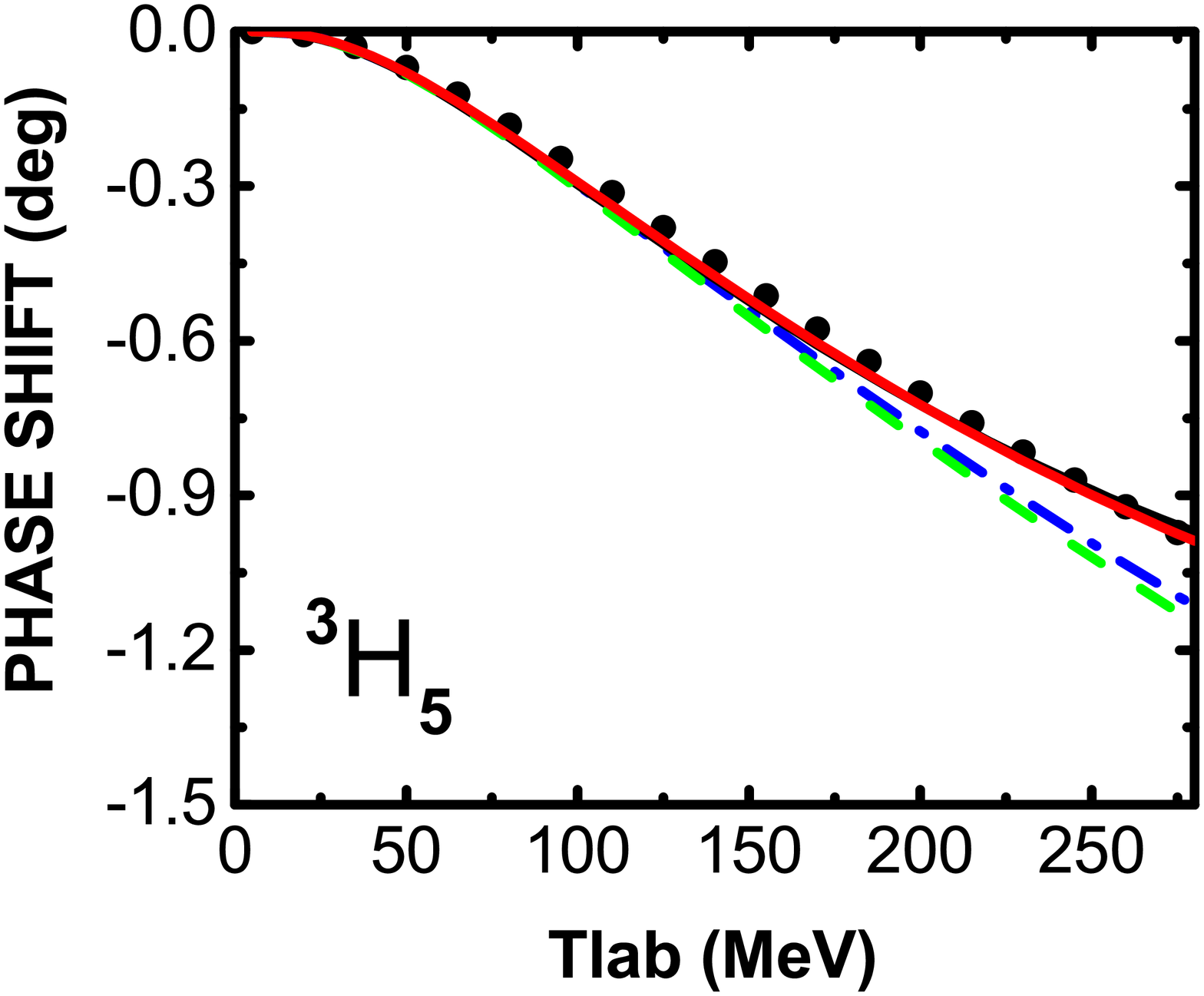}
}
\quad
\subfloat{
\includegraphics[width=0.45\textwidth]{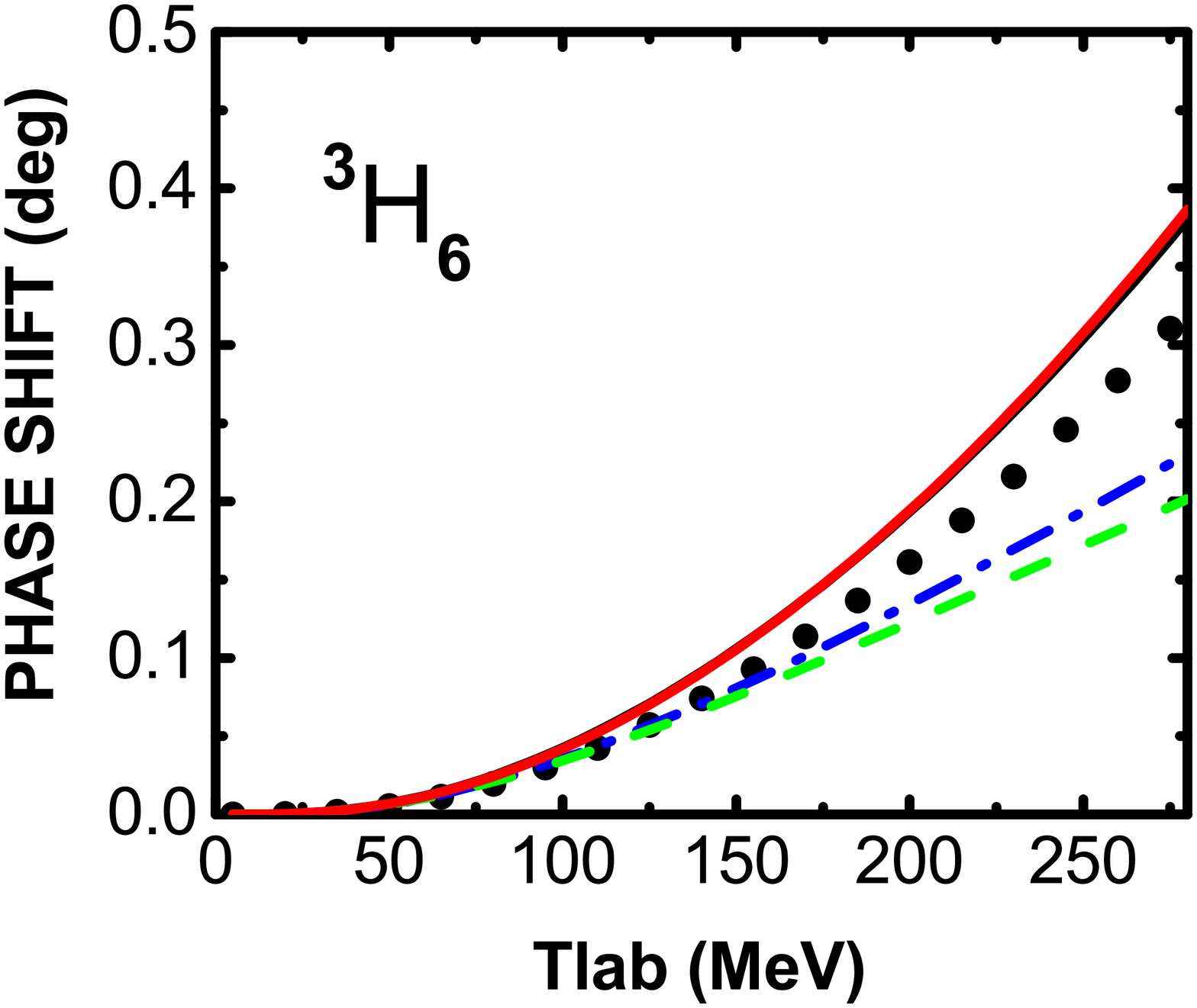}
}
\quad
\subfloat{
\includegraphics[width=0.45\textwidth]{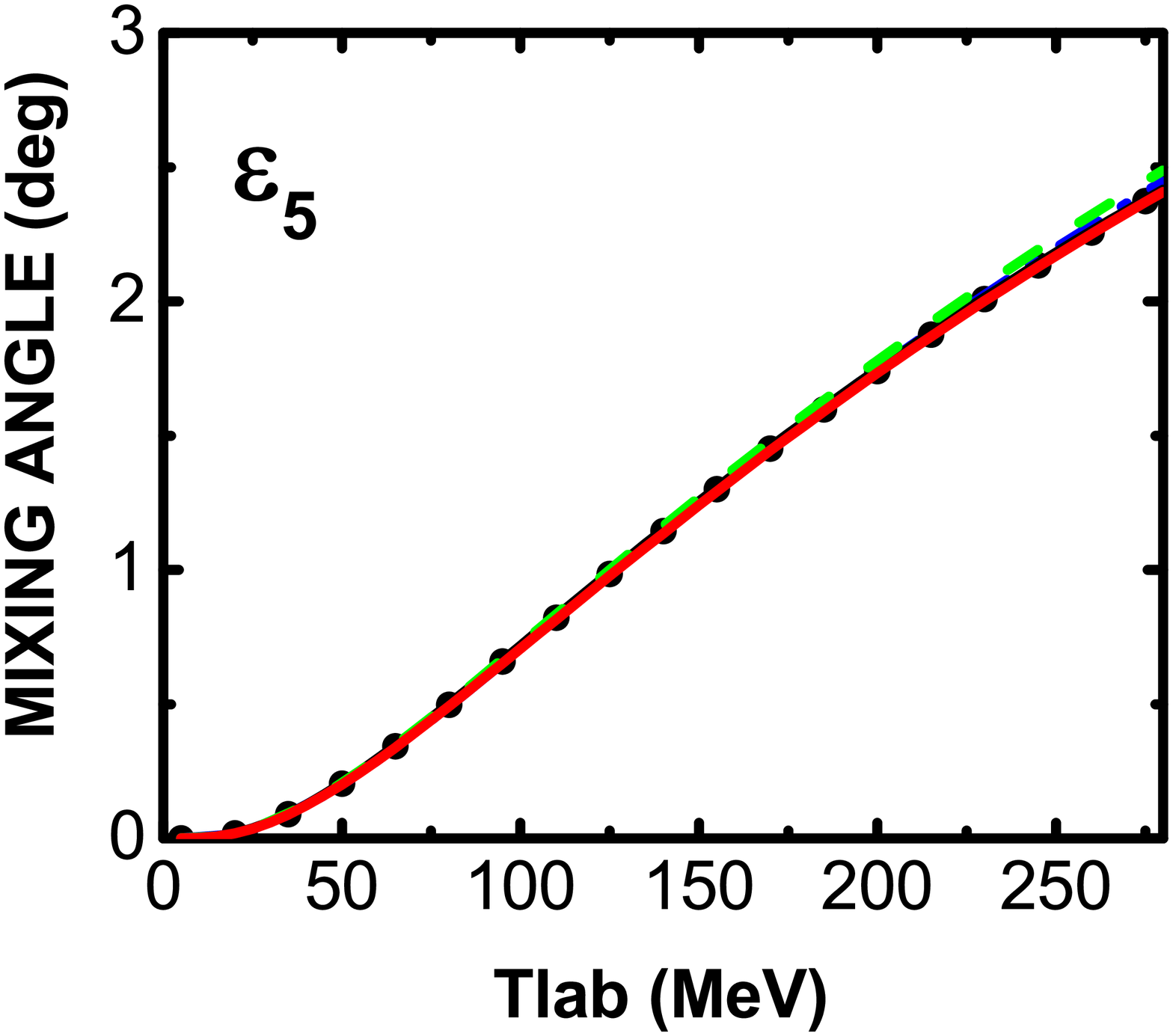}
}
\caption{Same as Fig.~\ref{tb:Dwave}, but for H-wave phase shifts and mixing angle $\epsilon_5$.}
\label{tb:Hwave}
\end{figure}

The H-wave phase shifts and mixing angle $\epsilon_5$ are depicted in Fig.~\ref{tb:Hwave}. For the H wave, {the results are independent of $\mu$ and} although the TPE contributions are much smaller, the relativistic corrections still improve the description of data. For the $^1\text{H}_5$ {and} $^3\text{H}_5$,  the relativistic and nonrelativistic phase shifts are almost indistinguishable. {For the $^3\text{H}_4$ partial wave, the relativistic results are slightly better}. Only for $^3\text{H}_6$, the contribution of the next-to-leading order TPE seems to be a bit large when $T_{\text{lab}}\geq150$ MeV. {Moreover, the contributions from the next-to-leading order TPE are still larger than those from the leading order for the $^1\text{H}_5$, $^3\text{H}_5$, $^3\text{H}_6$ partial waves and mixing angle $\epsilon_5$  as explained above.}

\subsection{I-wave}
\begin{figure}[htbp]
\centering
\subfloat{
\includegraphics[width=0.45\textwidth]{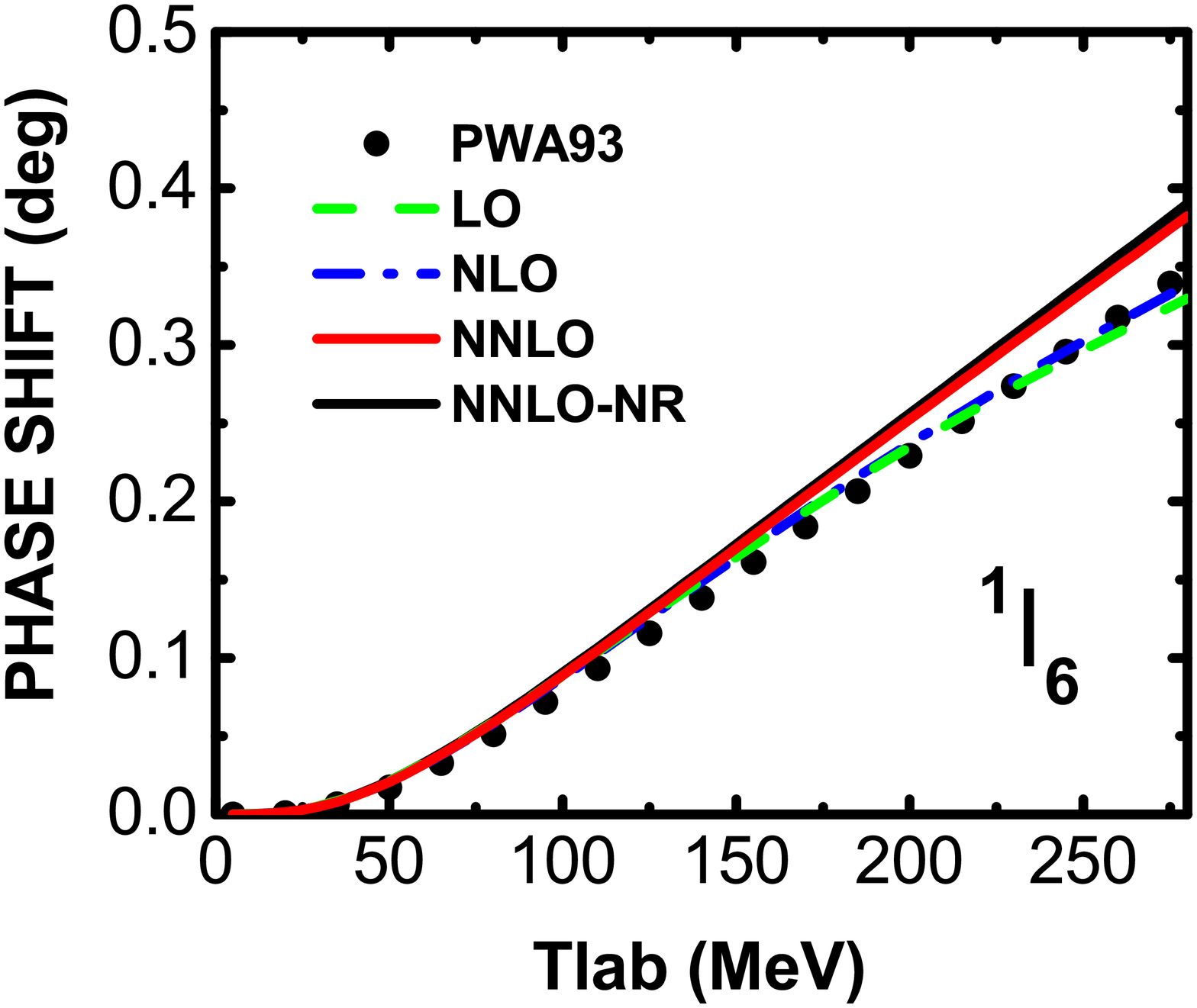}
}
\quad
\subfloat{
\includegraphics[width=0.45\textwidth]{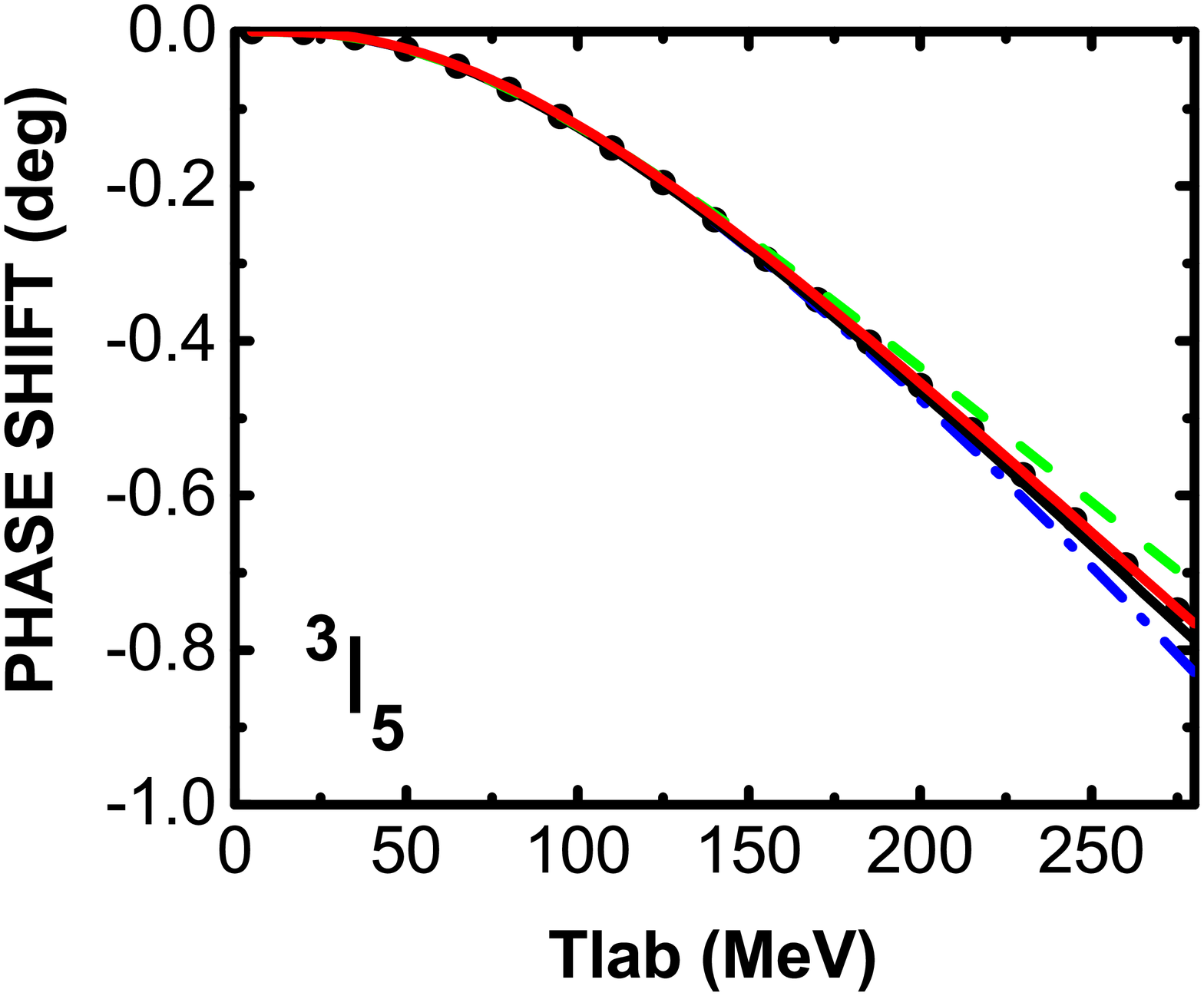}
}
\quad
\subfloat{
\includegraphics[width=0.45\textwidth]{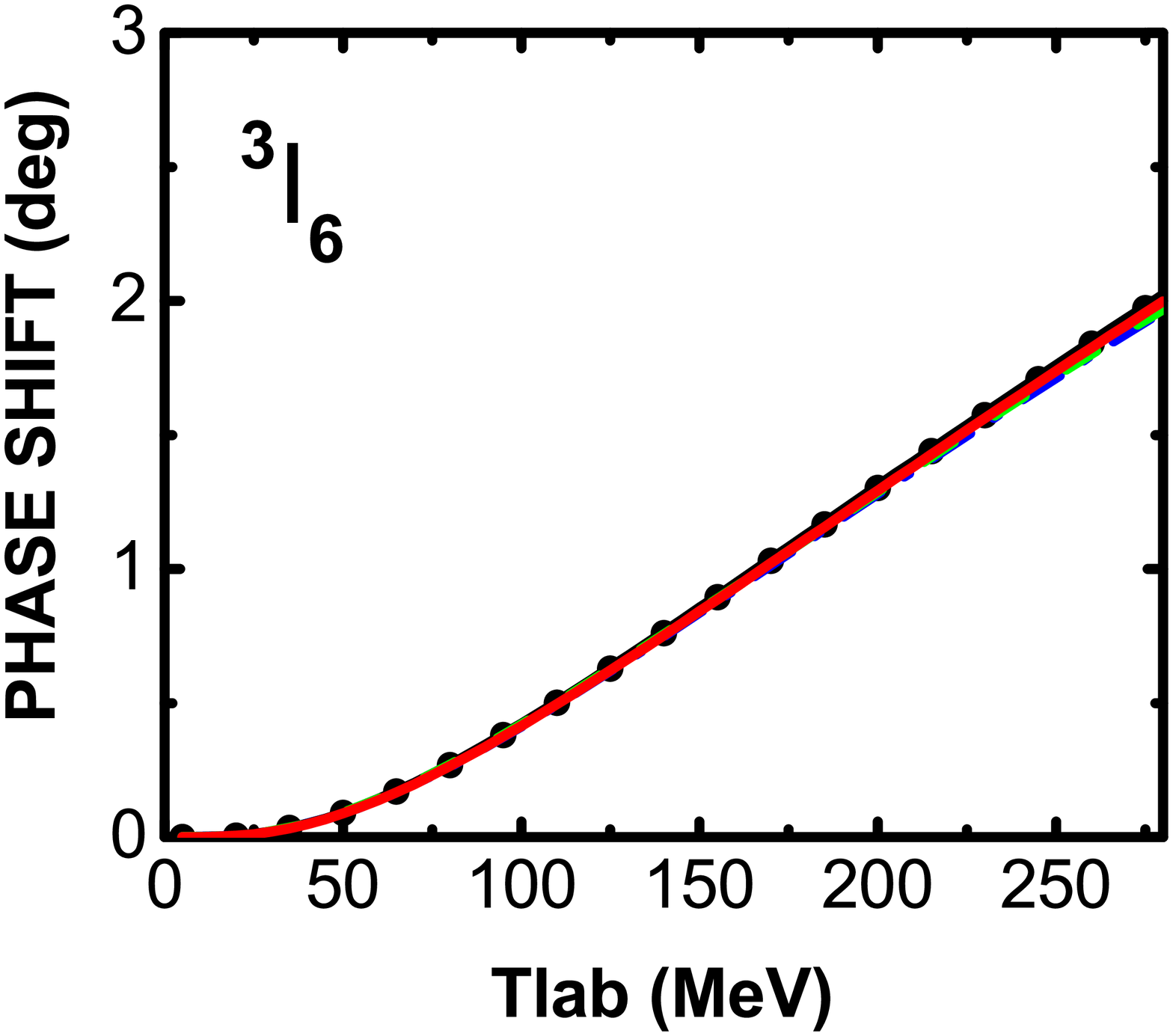}
}
\quad
\subfloat{
\includegraphics[width=0.45\textwidth]{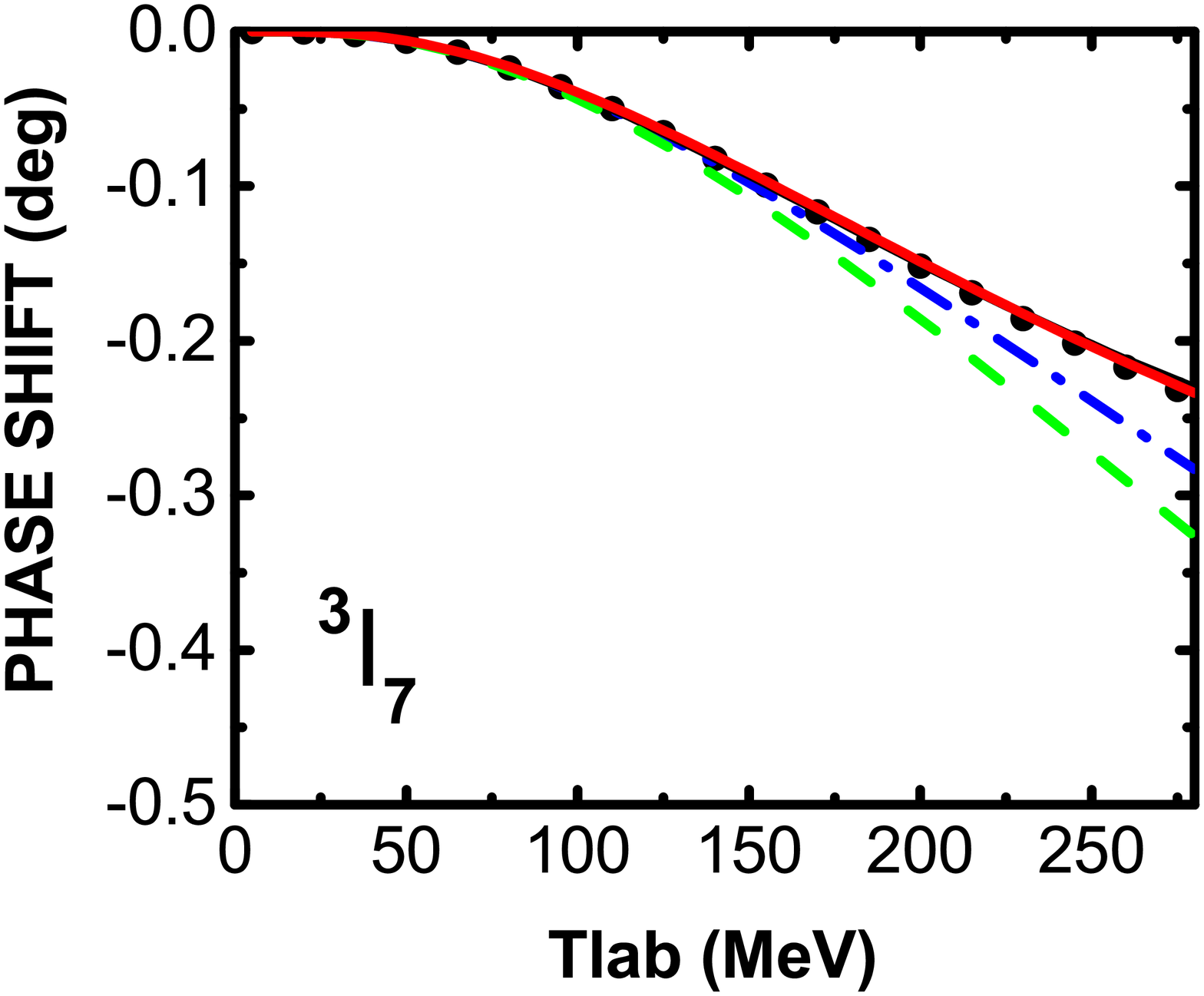}
}
\quad
\subfloat{
\includegraphics[width=0.45\textwidth]{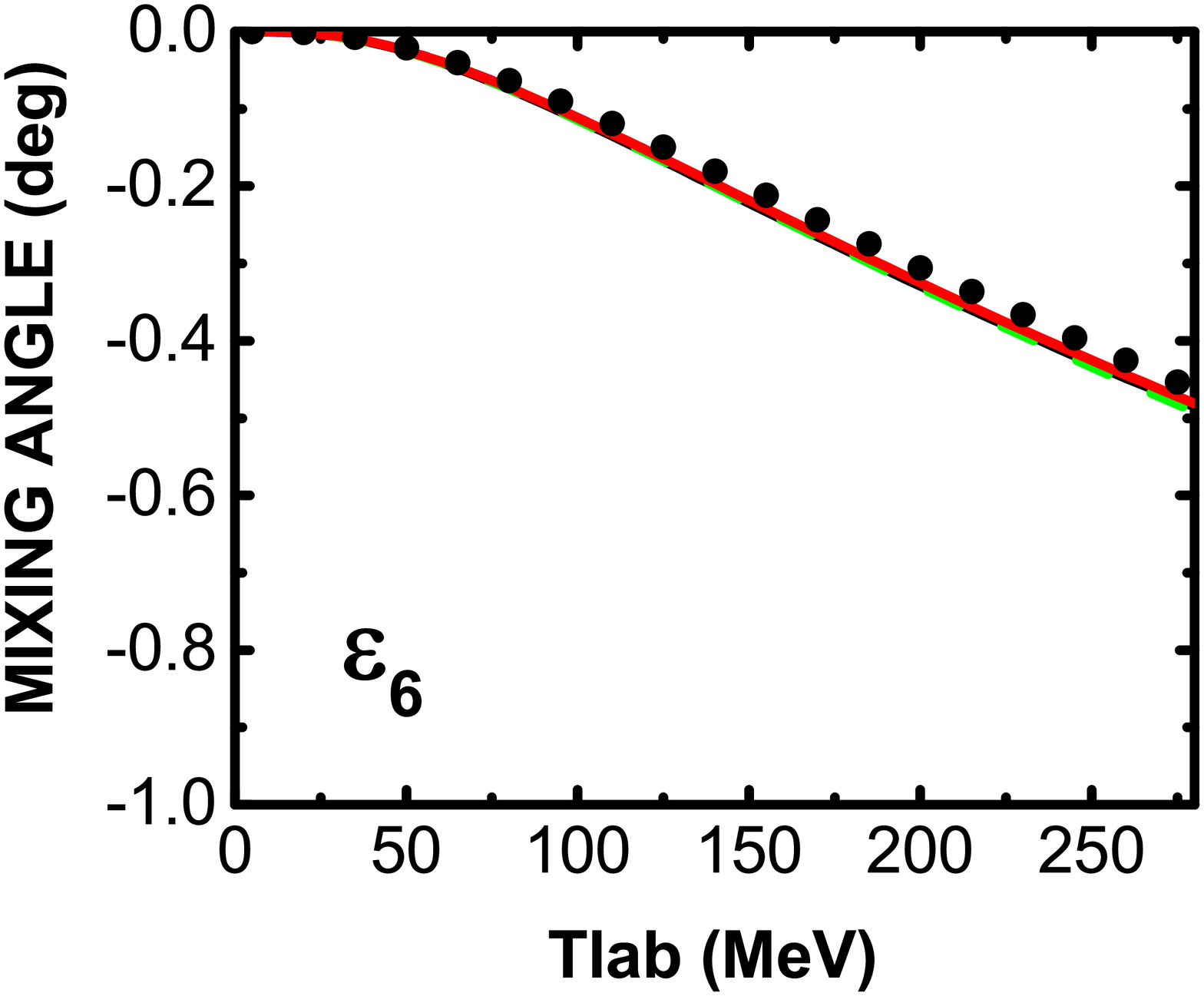}
}
\caption{Same as Fig.~\ref{tb:Dwave}, but for I-wave phase shifts and mixing angle $\epsilon_6$.}
\label{tb:Iwave}
\end{figure}

The I-wave phase shifts and mixing angle $\epsilon_6$ are depicted in Fig.~\ref{tb:Iwave}. {The results again are independent of $\mu$ and} the relativistic phase shifts are nearly identical to the nonrelativistic phase shifts and are in perfect agreement with data for this partial wave due to the negligible contribution of TPE. Notice that for the $^3\text{I}_7$ partial wave, the Nijmegen partial wave phase shifts~\cite{Stoks:1993tb} are larger than those in Ref.~\cite{Arndt:1986jb}. {As expected, the contributions from the next-to-leading order TPE are larger than those from the leading order for the $^1\text{I}_6$, $^3\text{I}_6$, $^1\text{I}_7$ partial waves and mixing angle $\epsilon_6$.}

\section{Summary and outlook}
Based on the covariant $\pi N$ Lagrangians, we calculated the relativistic TPE $T$-matrix up to $O(p^3)$. With this $T$-matrix, we further calculated the  chiral $NN$ phase shifts with $2\leq L\leq6$ and mixing angles with $2\leq J \leq6$ and then compared our results with those of the nonrelativistic expansion. We found that for all the partial waves the contributions of relativistic TPE are more moderate than their nonrelativistic counterparts and therefore the obtained $NN$ phase shifts are in better agreement with the Nijmegen partial wave analysis than the nonrelativistic results~\cite{Kaiser:1997mw} especially for the F partial waves.  Moreover, we showed that the large discrepancies between the nonrelativistic phase shifts and data in the $^3\text{F}_2$ partial wave can be eliminated by including the relativistic corrections. But for the $^3\text{F}_4$ partial wave, the relativistic corrections are insignificant. We found that the contributions of relativistic TPE at the next-to-leading order, similar to their nonrelativistic counterparts,
 are a bit large for the {$^1 J_J$, $^3 J_J$, $^3 (J-1)_J$} partial waves {and mixing angles} when $T_{\text{lab}}\geq150$ MeV {because of the large contributions from $c_3$ and $c_4$}, which indicates that the perturbation theory up to $\mathcal{O}(p^3)$ may not work well in this energy region.

 To summarize, although relativistic corrections are found to improve the description of data as expected, they are not significant enough to alter the results of Ref.~\cite{Kaiser:1997mw} at least at a qualitative level, thus
 supporting all the existing studies using the nonrelativistic two-pion exchange contributions of Ref.~\cite{Kaiser:1997mw} as inputs.
 On the other hand, given the covariant nature of the two-pion exchanges presented in this work, they can be easily utilized in the
 recent series of works~\cite{Ren:2016jna,Li:2016mln,Ren:2017yvw,Song:2018qqm,Li:2018tbt,Xiao:2018jot,Wang:2020myr,Bai:2020yml} which need such two-pion exchanges as inputs and their relevance in such settings remain to be explored.

\section{Acknowledgements}
Yang Xiao thanks Ubirajara L. van Kolck for useful discussions. This work is supported in part by the National
 Natural Science Foundation of China under Grants Nos.11735003, 11975041,  and 11961141004. Yang Xiao acknowledges the support from China Scholarship Council.


\end{document}